\documentclass[a4paper,pra,10pt,byrevtex,titlepage,secnumarabic,twocolumn,showpacs,showkeys,nobalancelastpage,nofootinbib]{revtex4-1}
\usepackage{graphicx}
\usepackage{amsmath}
\usepackage{amssymb}
\usepackage{natbib}
\usepackage{url}
\usepackage[squaren]{SIunits}
\usepackage[utf8]{inputenc}
\usepackage[spanish]{babel}
\usepackage{array}
\newcommand{\ucomer}{Comer}
\newcommand{\udormir}{Dormir}
\newcommand{\ucasa}{En Casa}
\newcommand{\utrabajar}{Trabajar}
\newcommand{\uhora}[1]{\mbox{#1}}
\decimalpoint
\addunit{\uday}{d}
\begin{document}
\author{José María Martín Olalla}
\email{olalla@us.es}
\homepage{@MartinOlalla\_JM}
\affiliation{Departamento de Física de la Materia Condensada. Universidad de Sevilla. Ap Correos 1065, ES41080. Sevilla}
\title{¿De verdad son tan anómalos los horarios españoles? Una comparación con Italia y Reino Unido}
\keywords{Huso horario; jetlag; trabajar; dormir; actividades; hora solar; hora civil; conciliación; uso del tiempo}
\published[Publicado:~]{16 de junio del 2014}
\begin{abstract}
  En el contexto de la discusión actual sobre el huso horario que debe regir en España este informe analiza las encuestas de uso de tiempo de España, Italia y Reino Unido y compara las estadísticas de actividades universales como comer o dormir de estos países en función de su longitud geográfica y de su latitud geográfica.

La comparación entre los datos de las regiones de estos países no se hace con las horas civiles ---ya que estas dependen del huso horario--- sino con la hora que mediría un observador extrínseco con una base de tiempo común y con la hora solar local. Hecha de esta forma la comparativa no muestra grandes anomalías entre las horas de las actividades de los tres conjuntos de datos. 

También se analiza la influencia de la latitud en el problema de la determinación de los horarios con una explicación plausible sobre por qué los horarios españoles parecen tan tempranos en el segmento matutino.

\end{abstract}

\maketitle

\renewcommand{\tocname}{Índice}
\tableofcontents
\makeatletter
\let\toc@pre\relax
\let\toc@post\relax
\makeatother 

\section{Motivaci{\'o}n}
\label{sec:motivacion}

En los últimos meses se observa en España una polémica intensa acerca los efectos perjudiciales que, sobre diferentes aspectos de la vida diaria, tienen los horarios españoles. Se señala que estos horarios no tienen parangón con ningún otro país de nuestro entorno y el culpable de que esto sea así es el huso horario que rige en España.\cite{bergareche-cambiar,fernandezcrehuet-fedea} Incluso se le identifica como una de las características del \emph{motto} \emph{Spain is different}.\cite{bolano-diferent} 

Desde el año 1940 este huso es el huso horario de la Europa central CET ---salvo en Canarias donde rige el huso horario de la Europa occidental WET. El huso CET está centrado sobre el meridiano $\unit{15}{\degree E}$ ---equivale a UTC+01\footnote{Las siglas UTC corresponden a \emph{tiempo universal coordinado} que es la forma correcta de referirse hoy al huso GMT ---tiempo medio de Greenwich---}--- alejado de la posición geográfica de la península ibérica por cuya parte oriental pasa el meridiano principal y el huso horario UTC+00.  El efecto primario de dicho desfase es que el mediodía no ocurre en ningún lugar de España a las 12h sino que lo hace entre cincuenta y noventa minutos después.\footnote{Cuando el horario de verano está vigente habría que añadir sesenta minutos adicionales. Técnicamente cabría decir con mayor propiedad que el horario español está centrado en el meridiano del horario de verano, $\unit{30}{\degree E}$, donde pasa siete meses del año. Solo en el horario de invierno (cinco meses) está centrado en los $\unit{15}{\degree E}$.} Estas críticas pueden resumirse en la siguiente cita:
\begin{quote}
  "<Vivimos en un \emph{jet lag} permanente. Como nuestra hora oficial no se corresponde con la hora solar, nuestras costumbres están alteradas."> (Nuria Chinchilla\footnote{Nuria Chinchilla era ese día directora del Centro Internacional Trabajo y Familia de la escuela de negocios IESE}  en Ref.~[\onlinecite{vidales-jetlag}])
\end{quote}

Pero, aunque aparentemente incontestable, la afirmación es errónea y el error está en la propia exposición: "<como nuestra hora oficial no se corresponde con la hora solar..."> no podemos comparar nuestros horarios con los horarios de lugares en los que sí coincide la hora oficial con la solar.

Nada resume mejor el error que se induce por este razonamiento que esta cita del \emph{Informe de la Subcomisión creada en el seno de la Comisión de Igualdad para el Estudio de la Racionalización de Horarios, la Conciliación de la Vida Personal, Familiar y Laboral y la Corresponsabilidad}:\cite{informe-subcomision}
\begin{quote}
  El problema de la conciliación en España se ve agravado además por una extraña organización diaria del horario que no tiene parangón con ningún otro país, y cuyo origen no parece encontrarse en razones geográficas, climatológicas o culturales, sino más bien en una curiosa circunstancia histórica, consistente en que durante la Segunda Guerra Mundial el Gobierno cambiase la hora española para adecuarse a la alemana, como hicieron otros países por razones bélicas ---Gran Bretaña, Francia, Portugal--- pero una vez terminada la guerra, no se volviese a la hora anteriormente en vigor, la del meridiano de Greenwich, cosa que sí hicieron Gran Bretaña y Portugal. De hecho España se encuentra ahora en el huso horario de Europa central, cuando debería estar en el huso europeo occidental. \emph{Por esta razón, nuestro horario se rige más por el sol que por el reloj: comemos a la una de la tarde, hora solar, y cenamos a las ocho, aunque nuestro reloj indique que son las tres y las diez de la noche, respectivamente\emph{[énfasis mío]}.} Esta situación se agrava más cuando adoptamos el horario de verano, cuando nuestra diferencia con respecto al sol es de tres horas. (página 74)
\end{quote}

Pasando por alto las imprecisiones históricas de la cita,\footnote{Portugal permaneció neutral en la Segunda Guerra Mundial y decir que el huso se cambió para "<adecuarlo"> con la "<hora alemana"> es una reducción: en 1938, la II República ya había empleado este horario\cite{planesas-ign-2013}; en 1940 el caso señalado de Portugal y Gran Bretaña: había razones para hacer el cambio distintas a adecuarse a la hora alemana. Francia adoptó de facto el horario alemán al ser ocupada durante la Segunda Guerra Mundial. Tras el fin de la Segunda Guerra Mundial Francia rehusó deshacer el cambio.} el objetivo de este trabajo es mostrar que, efectivamente, es el Sol sigue aún determinando los horarios; no solo en España ---con su "<huso incorrecto"> sino también en países donde rige el "<huso correcto">. Es decir, tratara de responder a cuestiones como: ¿en qué medida los horarios españoles son anómalos respecto de los horarios del resto de Europa? ¿cómo realizamos esa comparación?

Hace mucho tiempo leí una frase lapidaria en un artículo de la wikipedia inglesa que no puedo citar hoy pero que sí recuerdo: aunque en España rija el horario CET, los españoles hacen su vida en el horario WET ---el que le corresponde naturalmente por su meridiano. Es una idea que puede leerse de vez en cuando por aquí y por allí. Hasta hace poco no hacía más que repetir esta frase sin mucho éxito y sin datos que la avalaran. Pero recientemente leí un artículo\cite{albertos-amanece} en el que se señalaba, con cierto asombro, que las horas de despertar de los ciudadanos españoles dependían de la longitud. Mi sensación inicial es que eso era algo que debía ser natural toda vez que las actividades humanas están regidas por el Sol incluso en el siglo \textsc{XXI}. Inmediatamente después comprendí que, aún siendo obvio, merece la pena perder un tiempo en explicar esto porque la Ciencia, a veces, también tiene que ocuparse de explicar lo que es obvio.\footnote{Una idea atribuida a Proclo quien señaló que aunque un burro sepa que el camino más corto entre dos puntos es la línea recta, la Ciencia ---es decir, la geometría--- debe explicarlo.}

Gracias a ese artículo supe que existían las llamadas encuestas de Uso de Tiempo. El análisis de esta información  confirmó la vieja cita de la wikipedia. O dicho de otra forma: no hay nada malo en los horarios españoles; ninguna alteración significativa; ningún \emph{jet lag}; son similares ---dentro de un nivel razonable--- a los italianos o a los británicos. 

Para llegar a esta conclusión hay que geolocalizar los horarios. Esto quiere decir que la posición geográfica importa para dar sentido a la estructura horaria. Importa por la longitud geográfica ya que el Sol se mueve con una cadencia constante: cada hora barre quince grados de longitud de oriente a occidente. Pero también por la latitud: no pueden ser iguales los horarios de un lugar subtropical que los de un país cercano a un círculo polar.

Geolocalizar implica también estudiar los horarios desde una perspectiva diferente a la tabla horaria civil. Bien extrínseca, con un observador externo que registre los horarios con un reloj fijo. Bien intrínseca, analizando una propiedad que elimine la influencia del movimiento diario del Sol.  

En la sección~\ref{sec:resultados} se presentarán los resultados. Anteriormente en la sección~\ref{sec:metodologia} se describe la metodología con la que se ha realizado este estudio. En la sección~\ref{sec:influencia-de-la} se discute la influencia de la latitud en la determinación de los horarios de inicio de jornada, un aspecto del problema quizá menos conocido. Posteriormente se discuten otros temas relacionados con los horarios españoles.

\subsection{De qu{\'e} no trata este estudio}
\label{sec:de-que-no}

Este estudio no trata otros aspectos de la polémica actual: por ejemplo si la jornada laboral española ha de tener tal o cual estructura; si los ciudadanos españoles están mucho o poco pluriempleados, más o menos que los ciudadanos italianos; o en qué medida podemos mejorar la conciliación de la vida laboral y familiar. Estos temas organizativos son fundamentalmente independiente del huso que esté vigente en un país. Hagamos lo que hagamos con el huso el día va a seguir teniendo veinticuatro horas y el Sol va a seguir saliendo y poniéndose cuando le toca.

Tampoco se trata en este estudio del asunto del horario de verano. Aunque se suele decir que el horario de verano se implementa para ahorrar energía desde un punto de vista aséptico la cuestión se resume con la siguiente idea: el cambio de hora de verano hace que el amanecer ocurra en un rango de tiempo estrecho, menor que el que ocurriría si no se hiciera el cambio. Es decir, se hace el cambio para que la hora del orto solar no varíe demasiado a lo largo del año: el precio es la enorme variación de la hora del ocaso.

Utilizaré todos los valores que aparecen en la encuesta de tiempo que incluyen respuestas referidas a meses en los que rige el horario de verano y respuestas referidas a encuestas en las que rige el horario de invierno. No distinguiré estas situaciones y lo que se obtiene ha de ser un comportamiento promedio anual. Además cuando me refiera a un huso lo haré normalmente asumiendo que es el huso del horario de invierno. Así si se dice que Italia vive en CET nos estamos refiriendo a su horario de invierno. En verano se convierte en CEST ---horario de verano de la Europa central---.

\section{Metodolog{\'\i}a}
\label{sec:metodologia}

\subsection{Aspectos generales}
\label{sec:aspectos-generales}

En este informe se analizan las microdatos de las encuestas de uso del tiempo que se listan en el cuadro~\ref{tab:encuestas}.

\begin{table*}[tb]
  \centering
  \begin{tabular}{llllp{4cm}}
\toprule
    \textbf{Institución}&\textbf{País}&\textbf{Encuesta}&\textbf{Año}&\textbf{Detalle regional}\\
\colrule
    Instituto Nacional de Estadística&España&Empleo del Tiempo&2009-2010&comunidades y ciudades autónomas (17+1)\\
    Istituto Nazionale di Statistica, Istat&Italia&Uso del tempo&2008-2009&regione (19)\\
    UK Data Service&Reino Unido&Time Use Survey&2000&regions of England y resto de \emph{home-nations} (9+3)\\
\botrule
  \end{tabular}
  \caption{Identificación de las encuestas analizadas en este informe. En la última columna aparece el nivel de parcelación regional, entre paréntesis el número de regiones.}
  \label{tab:encuestas}
\end{table*}

Las encuestas de uso del tiempo se realizan con una periodicidad variable en un gran número de países. Aparte de otros muchos aspectos que caracterizan al individuo encuestado ---edad, sexo, ingresos o situación laboral--- la encuesta registra el día de la semana al que se refiere el diario de actividades, lo que permite tratar de forma diferente los días laborales (de lunes a viernes) del fin de semana: en este estudio solo se analizan días laborales. 

Las encuestas están parceladas geográficamente por regiones administrativas tal y como aparece en el cuadro~\ref{tab:encuestas} siendo imposible conocer una localización más precisa del dónde se sitúa el hogar del entrevistado. En este estudio los datos de las encuestas se analizan agrupados por esas regiones: en ningún caso se calculan o analizan valores a nivel organizativo superior. Se admite pues como hipótesis de trabajo que los datos referidos a una región  están correctamente segmentados por municipio, sexo, edad, condición laboral etcétera, y son representativos de la región.

 Las tres encuestas analizadas compartimentan el día en fracciones de diez minutos de forma que hay seis registros por hora y ciento cuarenta y cuatro registros por día. Las encuestas de Italia y el Reino Unido se inician a las \uhora{04:00} y terminan un día después a la misma hora. La encuesta de España hace lo mismo para las \uhora{06:00}.  Inevitablemente los datos obtenidos tienen una discreción notable. En parte es reflejo del diseño de la encuesta pero también el lógico pensar que las personas programan actividades (típicamente el despertar) en valores exactos de horas. Incluso si no fuera así, las personas solo recuerdan la hora de realización de actividades con una precisión de diez minutos o un cuarto de hora. 

Se va a analizar las dos actividades más universales ---"<\ucomer"> y "<\udormir">--- y la localización más universal ---"<\ucasa">---. Además se analizará la actividad "<\utrabajar">, cuya frecuencia es sustancialmente menor. De estos cuatro estudios tres son puramente booleanos. "<\udormir">, "<\utrabajar"> o "<\ucasa"> solo pueden tener dos estados posibles: cierto o falso. 

La codificación de la actividad "<\ucomer"> es más compleja ya que la ingesta de alimentos es, también, un problema cuantitativo. En el elenco de actividades de la encuesta italiana  aparece "<(211) pasti principali"> que se refiere a comidas fuertes. En las encuestas española e inglesa aparece solo "<(21) comer (eating)"> válido para cualquier ingesta de alimentos por grande o pequeña que sea.

Para cada actividad o localización analizada se ha determinado los tiempos de inicio y fin más tempranos en el día y más tardíos en el día. Temprano o tardío se define en relación con la hora de inicio de la encuesta que son la cuatro de la mañana en Italia y Reino Unido y las seis de la mañana en España. %

La forma en la que se determinan las ocurrencias más tempranas y tardías implica que  una hora de despertar de  \uhora{03:30} ---es decir, $\unit{3.5}{\hour}$ después del inicio del día civil--- será un valor relativamente tardío ya que han pasado $\unit{23.5}{\hour}$ desde las cuatro de la mañana (inicio de la actividad en Italia y Reino Unido) o $\unit{21.5}{\hour}$ desde la seis de la mañana (inicio de la actividad en España). De la misma forma, volver a casa a las \uhora{03:30} es volver tarde, pero volver a las \uhora{07:00} es volver temprano.

Se toma un cuidado especial en analizar la actividad que se repite al principio y final del registro. Típicamente dicha actividad es la de dormir.

Dependiendo de la actividad analizada los datos más tardíos y más tempranos de horas de inicio y fin de la actividad adquieren un significado; en el cuadro~\ref{tab:casos} se lista la identificación de estas variables con sus equivalencias genéricas. En la actividad "<\udormir">, la hora más temprana de finalización la asociamos al despertar; la hora más tardía de inicio la asociamos a acostarse. De la misma forma, para la localización "<\ucasa">, la hora más temprana de finalización es la hora de salida del hogar y la hora más tardía de inicio es la hora de regreso al hogar. Obviamente en todos estos casos hay una simplificación estadística implícita ya que en ambos casos es solo la primera (última) ocurrencia de la actividad.

\begin{table*}[tb]
  \centering
  \begin{tabular}{lll}
\toprule
    \textbf{Actividad/Localización}&\textbf{Variable determinada}&\textbf{Significado}\\
\colrule
    \udormir&Hora de finalización más temprana&Despertarse\\
    \udormir&Hora de inicio más tardío&Acostarse\\
    \udormir&Tiempo acumulado&Tiempo de sueño\\
    \ucomer&Hora de inicio más temprano&Desayunarse\\
    \ucomer&Hora de inicio más tardío&Cena\\
    \ucomer&Tiempo acumulado&Tiempo comiendo\\
    \ucasa&Hora de fin más temprano&Salir del hogar\\
    \ucasa&Hora de inicio más tardío&Regresar al hogar\\
    \ucasa&Tiempo acumulado&Tiempo en el hogar\\
    \utrabajar&Hora de inicio más temprano&Entrar al trabajo\\
    \utrabajar&Hora de fin más tardío&Salir del trabajo\\
   \utrabajar&Tiempo acumulado&Tiempo trabajando\\
\botrule
  \end{tabular}
  \caption{Variables determinadas en el registro de actividades de las encuesta de tiempo y equivalencia para cada una de las actividades o localizaciones estudiadas.}
  \label{tab:casos}
\end{table*}

En el cuadro~\ref{tab:numero} se listan el número de respuestas analizadas para cada actividad-localización y cada región analizada y en  días laborables. Solo entran en el cómputo las personas que realizan la actividad en algún momento del día.

\begin{table*}
  \centering
  \begin{tabular}{rlllrrrr}
\toprule
&&&&\textbf{\udormir}&\textbf{\ucomer}&\textbf{\ucasa}&\textbf{\utrabajar}\\
    \textbf{Código}&\textbf{Región}&\textbf{Longitud}&\textbf{Latitud}&$N_1$&$N_2$&$N_3$&$N_4$\\
\colrule
\multicolumn{8}{c}{\textbf{España}}\\
$1$&Andalucía&$\unit{-4.87}{\degree}$&$\unit{+37.1}{\degree}$&$ 1322$&$ 1321$&$ 1210$&$ 401$\\
$2$&Aragón&$\unit{-0.807}{\degree}$&$\unit{+41.6}{\degree}$&$ 476$&$ 476$&$ 450$&$ 194$\\
$3$&Principado de Asturias&$\unit{-5.82}{\degree}$&$\unit{+43.4}{\degree}$&$ 442$&$ 442$&$ 409$&$ 147$\\
$4$&Illes Balears&$\unit{+2.73}{\degree}$&$\unit{+39.5}{\degree}$&$ 438$&$ 438$&$ 423$&$ 179$\\
$5$&Canarias&$\unit{-15.8}{\degree}$&$\unit{+28.3}{\degree}$&$ 461$&$ 461$&$ 419$&$ 126$\\
$6$&Cantabria&$\unit{-3.82}{\degree}$&$\unit{+43.4}{\degree}$&$ 409$&$ 409$&$ 382$&$ 159$\\
$7$&Castilla y León&$\unit{-4.86}{\degree}$&$\unit{+41.8}{\degree}$&$ 641$&$ 641$&$ 610$&$ 223$\\
$8$&Castilla-La Mancha&$\unit{-3.21}{\degree}$&$\unit{+39.5}{\degree}$&$ 604$&$ 604$&$ 573$&$ 230$\\
$9$&Cataluña&$\unit{+2.04}{\degree}$&$\unit{+41.5}{\degree}$&$ 1028$&$ 1028$&$ 979$&$ 473$\\
$10$&Comunidad Valenciana&$\unit{-0.415}{\degree}$&$\unit{+39.1}{\degree}$&$ 778$&$ 777$&$ 729$&$ 280$\\
$11$&Extremadura&$\unit{-6.34}{\degree}$&$\unit{+39.1}{\degree}$&$ 450$&$ 450$&$ 421$&$ 155$\\
$12$&Galicia&$\unit{-8.34}{\degree}$&$\unit{+42.8}{\degree}$&$ 836$&$ 836$&$ 760$&$ 324$\\
$13$&Comunidad de Madrid&$\unit{-3.71}{\degree}$&$\unit{+40.4}{\degree}$&$ 1404$&$ 1404$&$ 1322$&$ 617$\\
$14$&Región de Murcia&$\unit{-1.22}{\degree}$&$\unit{+37.9}{\degree}$&$ 374$&$ 374$&$ 351$&$ 130$\\
$15$&Comunidad Foral de Navarra&$\unit{-1.68}{\degree}$&$\unit{+42.7}{\degree}$&$ 829$&$ 830$&$ 801$&$ 329$\\
$16$&País Vasco&$\unit{-2.62}{\degree}$&$\unit{+43.2}{\degree}$&$ 430$&$ 430$&$ 408$&$ 155$\\
$17$&La Rioja&$\unit{-2.37}{\degree}$&$\unit{+42.4}{\degree}$&$ 388$&$ 387$&$ 378$&$ 156$\\
$18$&Ceuta y Melilla&$\unit{-4.17}{\degree}$&$\unit{+35.6}{\degree}$&$ 422$&$ 424$&$ 391$&$ 138$\\

\multicolumn{8}{c}{\textbf{Italia}}\\
$1$&Piemonte-Vall d'Aosta&$\unit{+7.88}{\degree}$&$\unit{+45.1}{\degree}$&$ 1467$&$ 1464$&$ 1451$&$ 608$\\
$2$&Lombardia&$\unit{+9.41}{\degree}$&$\unit{+45.6}{\degree}$&$ 1248$&$ 1232$&$ 1228$&$ 531$\\
$3$&Trentino Alto-Adige&$\unit{+11.3}{\degree}$&$\unit{+46.3}{\degree}$&$ 840$&$ 831$&$ 812$&$ 375$\\
$4$&Veneto&$\unit{+11.8}{\degree}$&$\unit{+45.5}{\degree}$&$ 837$&$ 830$&$ 818$&$ 345$\\
$5$&Friuli-Venezia Giulia&$\unit{+13.2}{\degree}$&$\unit{+45.9}{\degree}$&$ 528$&$ 520$&$ 523$&$ 196$\\
$6$&Liguria&$\unit{+8.86}{\degree}$&$\unit{+44.3}{\degree}$&$ 652$&$ 644$&$ 642$&$ 235$\\
$7$&Emilia Romagna&$\unit{+11.2}{\degree}$&$\unit{+44.6}{\degree}$&$ 799$&$ 790$&$ 786$&$ 319$\\
$8$&Toscana&$\unit{+10.9}{\degree}$&$\unit{+43.7}{\degree}$&$ 857$&$ 848$&$ 845$&$ 360$\\
$9$&Umbria&$\unit{+12.5}{\degree}$&$\unit{+43.0}{\degree}$&$ 480$&$ 474$&$ 476$&$ 193$\\
$10$&Marche&$\unit{+13.4}{\degree}$&$\unit{+43.4}{\degree}$&$ 650$&$ 647$&$ 640$&$ 265$\\
$11$&Lazio&$\unit{+12.6}{\degree}$&$\unit{+41.9}{\degree}$&$ 730$&$ 722$&$ 716$&$ 315$\\
$12$&Abruzzo&$\unit{+14.0}{\degree}$&$\unit{+42.4}{\degree}$&$ 551$&$ 546$&$ 545$&$ 218$\\
$13$&Molise&$\unit{+14.7}{\degree}$&$\unit{+41.7}{\degree}$&$ 461$&$ 460$&$ 457$&$ 139$\\
$14$&Campania&$\unit{+14.4}{\degree}$&$\unit{+40.9}{\degree}$&$ 981$&$ 976$&$ 974$&$ 345$\\
$15$&Puglia&$\unit{+17.0}{\degree}$&$\unit{+40.9}{\degree}$&$ 812$&$ 806$&$ 807$&$ 238$\\
$16$&Basilicata&$\unit{+16.1}{\degree}$&$\unit{+40.6}{\degree}$&$ 478$&$ 476$&$ 476$&$ 159$\\
$17$&Calabria&$\unit{+16.3}{\degree}$&$\unit{+38.9}{\degree}$&$ 723$&$ 718$&$ 711$&$ 238$\\
$18$&Sicilia&$\unit{+14.2}{\degree}$&$\unit{+37.7}{\degree}$&$ 1045$&$ 1035$&$ 1043$&$ 318$\\
$19$&Sardegna&$\unit{+8.96}{\degree}$&$\unit{+39.9}{\degree}$&$ 648$&$ 642$&$ 641$&$ 223$\\

\multicolumn{8}{c}{\textbf{Reino Unido}}\\
$1$&North East&$\unit{-1.47}{\degree}$&$\unit{+54.8}{\degree}$&$ 448$&$ 435$&$ 446$&$ 150$\\
$2$&North West&$\unit{-2.47}{\degree}$&$\unit{+53.4}{\degree}$&$ 1171$&$ 1144$&$ 1153$&$ 464$\\
$3$&Yorkshire and the Humber&$\unit{-1.28}{\degree}$&$\unit{+53.7}{\degree}$&$ 1035$&$ 1009$&$ 1020$&$ 455$\\
$4$&East Midlands&$\unit{-1.21}{\degree}$&$\unit{+52.8}{\degree}$&$ 960$&$ 941$&$ 945$&$ 408$\\
$5$&West Midlands&$\unit{-1.65}{\degree}$&$\unit{+52.4}{\degree}$&$ 807$&$ 792$&$ 778$&$ 313$\\
$6$&Eastern&$\unit{+0.478}{\degree}$&$\unit{+52.0}{\degree}$&$ 1073$&$ 1040$&$ 1053$&$ 458$\\
$7$&London&$\unit{-0.108}{\degree}$&$\unit{+51.5}{\degree}$&$ 830$&$ 810$&$ 810$&$ 351$\\
$8$&South East&$\unit{-0.459}{\degree}$&$\unit{+51.2}{\degree}$&$ 1344$&$ 1317$&$ 1322$&$ 564$\\
$9$&South West&$\unit{-2.80}{\degree}$&$\unit{+51.1}{\degree}$&$ 974$&$ 958$&$ 959$&$ 409$\\
$10$&Wales&$\unit{-3.43}{\degree}$&$\unit{+51.9}{\degree}$&$ 494$&$ 485$&$ 491$&$ 186$\\
$11$&Scotland&$\unit{-3.73}{\degree}$&$\unit{+56.1}{\degree}$&$ 1080$&$ 1048$&$ 1055$&$ 438$\\
$12$&Northern Ireland&$\unit{-6.26}{\degree}$&$\unit{+54.7}{\degree}$&$ 261$&$ 255$&$ 260$&$ 119$\\

\botrule
  \end{tabular}
  \caption{En la primera columna se listan los códigos numéricos que aparecen en la figura~\ref{fig:mapa} y que corresponden a las regiones cuyo nombre se lista en la segunda columna. A continuación sus valores promedios de longitud geográfica y latitud ---valores negativos de longitud significan al oeste del meridiano principal, valores positivos significa al este del meridiano principal; los valores positivos de latitud significan al norte del ecuador---. Finalmente el número de respuestas analizadas en cada una de las tres actividades y de la localización listadas en el cuadro~\ref{tab:casos} y para días laborables. Solo se analizan individuos el tiempo acumulado de actividad es distinto de cero.}
  \label{tab:numero}
\end{table*}

\subsection{Tres relojes: hora local civil, hora de Londres, hora local solar}
\label{sec:simulacion-de-la}

Jules Verne publicó "<Le tour de monde en quatre-vingts jours"> (\emph{La vuelta al mundo en ochenta días}) en 1873, allá cuando los viajes empezaban a tener la intensidad adecuada como para notar los efectos del cambio de meridiano. En esa época la base de tiempo era local: las 12:00 coincidían con el tránsito del Sol.\footnote{El tránsito solar es el momento en el que el Sol cruza el meridiano local en su movimiento aparente de revolución alrededor de la Tierra. Divide el recorrido del Sol en dos mitades: anterior ---\emph {ante meridiem (AM)}--- y posterior ---\emph{post meridiem (PM)}---} En la novela los dos personajes principales guardan el tiempo de forma diferente. Phileas Fogg adelanta escrupulosamente su reloj a razón de cuatro minutos por cada grado de meridiano avanzado hacia oriente. Passepartout, más olvidadizo, guarda en su reloj la hora de Londres.

 Poco después de la publicación de su novela se generalizó el sistema de huso horarios vigente en la actualidad. Si Verne publicara una versión actual de su novel podría considerar la posibilidad de añadir un personaje que sincronizase su reloj con la hora civil (el huso civil) vigente en el lugar y que varía discretamente de hora en hora.\footnote{En algunas pocas regiones del mundo el huso civil está desplazado media hora ---por ejemplo en Venezuela, India o Irán--- o un cuarto de hora ---caso de Nepal--- respecto de los husos horarios normales.}

Los datos de la encuesta de tiempo se toman y se publican en el huso civil vigente; es decir en el tiempo de este hipotético tercer personaje. Sin embargo, para hacer un estudio comparativo de los tiempos de estas encuestas es necesario considerar los otros dos relojes: el de Passepartout (reloj constante con una hora fija) y el de Fogg (el reloj solar local).

El hora del reloj de Passepartout se determina fácilmente sabiendo el huso local vigente y restando las horas adicionalmente añadidas. En el caso del estudio que vamos a realizar solo hay que restar una hora a las regiones de Italia y a todas las regiones de España salvo Canarias. Ésta y las regiones del Reino Unido marcan la hora de Londres en su hora civil.\footnote{Hay un problema adicional en la serie de datos cual es el uso del horario de verano. Podría analizarse las respuestas habidas en los meses de invierno y tratarlas de forma diferente a las respuesta de los veranos pero se va hacer un análisis global de todo el año. También por este motivo se usará el término "<hora de Londres"> queriendo significar huso WET o WEST ---horario de verano de la Europa Occidental--- según la época del año.} Matemáticamente se expresa con la fórmula:
\newcommand{\umod}{\ensuremath{\mathrm{mod}}}

\begin{equation}
  \label{eq:2}
  \tau_i=\left(t_i-\Delta_{z_i}+A\right)\umod A,
\end{equation}
donde $t_i$ es el tiempo que aparecen en la encuesta de uso de tiempo, $\Delta_{z_i}$ es la diferencia de la hora local con la hora de Londres y $\tau_i$ es el tiempo expresado en la hora de Londres. $A$  representa la duración de un día en las unidades en las que esté expresado el tiempo.\footnote{Típicamente $\unit{24}{\hour}$ pero en la encuesta de uso del tiempo los valores de tiempos se presenten en unidades de diez minutos. Hay seis de estas unidades por hora y ciento cuarenta y cuatro por día.} La operación módulo asegura que el resultado de la operación está en el intervalo $[0,A)$.\footnote{Por ejemplo un valor de $00:00$ en el huso UTC+01 ($t_z=\unit{1}{\hour}$) lo convierte en $23:00$ y no en $-01:00$.}

La hora de solar hay que computarla siguiendo la razón constante de cuatro minutos por grado de diferencia de longitud que utilizaba Phileas Fogg y que se representará por la letra griega $\nu$ en este informe. Esta constante multiplicada por la longitud geográfica del lugar expresa el desfase entre el mediodía local y el mediodía en Londres. Este desfase $\lambda\nu$ es avance si nos movemos hacia el este ---se toma $\lambda>0$--- y retardo en caso contrario ---se toma $\lambda<0$---. El desfase añadido a la  hora de Londres $\tau_i$ determina la hora solar a la que ocurre un evento. Esta hora se expresa entonces como:
\begin{equation}
  \label{eq:1}
  T_i=\left(\tau_i+\lambda_i\nu+A\right)\umod\, A=\left(t_i-\Delta_{z_i}+\lambda_i\nu+A\right)\umod A,
\end{equation}
donde $\lambda_i$ es la longitud geográfica del hogar del respondiente $i$.

La gran diferencia entre la hora solar y la local o la UTC es que las dos segundas miden la distancia temporal entre un evento y el tránsito solar por el punto antipodal al meridiano principal ---medianoche UTC--- o a cualquiera de sus meridianos horarios secundarios. La hora solar computa lo mismo pero en relación con el meridiano antipodal al meridiano local de observación.

Como la longitud de la ecuación~(\ref{eq:1}) no es conocida, se representará  por la longitud promedio de la región en la que esté inscrito el hogar. Este valor se determina promediando, según la población, la longitud (y también la latitud) de los municipios de más de mil habitantes de cada región:
\begin{equation}
  \label{eq:3}
  \langle\lambda\rangle_k=\frac{\sum_{j}P^k_j\lambda^k_j}{\sum_j P^k_j}
\end{equation}
donde $j$ son las poblaciones con más de mil habitantes de la región $k$, $\lambda_j^k$ es su longitud geográfica (o su latitud), $P_j^k$ es la población de la localidad y $\langle\lambda\rangle_k$ es el valor promedio ponderado de la longitud geográfica (o de la latitud) de la región $k$.

\begin{figure}
  \centering
  \includegraphics[bb= 108 63 328 287]{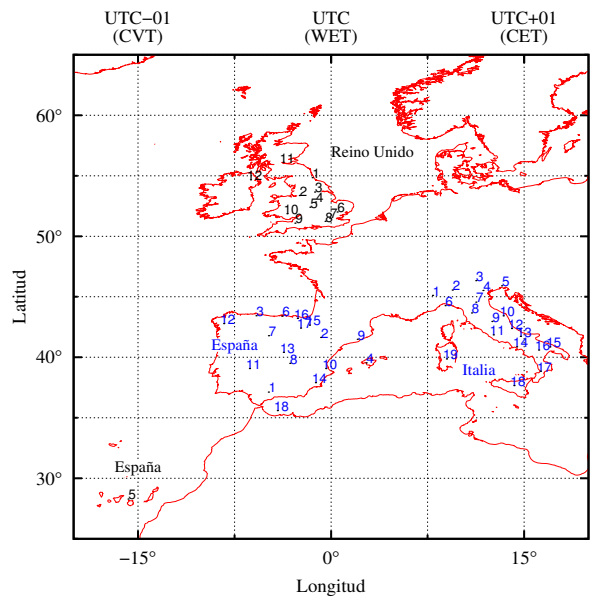}
  \caption{Distribución regional de España, Italia y Reino Unido. La etiqueta numérica identifica la región en el cuadro~\ref{tab:numero}. El punto identifica el valor promedio ponderado de la latitud y la longitud de la región. En el eje X superior se muestra los husos UTC correspondientes y sus etiquetas habituales: de occidente a oriente CVT (huso horario de Cabo Verde), WET (huso horario de Europa Occidental) y CET (huso horario de Europa Central). Las regiones cuya hora civil invernal se rige por el huso WET se muestran en negro, las regiones cuya hora civil invernal se rige por el huso CET se muestran en azul.}
  \label{fig:mapa}
\end{figure}

En la figura~\ref{fig:mapa} se muestra la localización geográfica de las regiones analizadas. El punto que aparece en la figura se representa en el valor promedio de la latitud y longitud de la región. La etiqueta identifica la región en el cuadro~\ref{tab:numero}. En la figura~\ref{fig:mapa} se muestra en azul las regiones con horario CET y en negro las regiones con horario WET.

La elección de los dos países es venturosa. Italia coincide ampliamente con España en latitud; está desplazada en longitud geográfica pero comparten el mismo huso civil. Reino Unido está situada geográficamente sobre la misma longitud de forma que el tránsito solar ocurre contemporáneamente con el tránsito solar en la península. Sin embargo observa un huso horario diferente y está situada más al norte.

La población de los tres países es parecida en de órdenes de magnitud. También la amplitud de la longitud geográfica (si no incluimos a Canarias en el cómputo de España). Además los tres países están cortados por un meridiano horario en su parte más oriental quedando la mayor parte de la población y del territorio al oeste de este meridiano.

\subsection{T{\'e}cnicas matem{\'a}ticas}
\label{sec:tecnicas-matematicas}

El análisis matemático realizado en este estudio consta de dos partes diferentes y complementarias.

Por una parte se construye la probabilidad acumulada empírica de la magnitud observada. Esta probabilidad acumulada expresa la fracción de población que cumple un determinado requerimiento a una determinada hora. Por ejemplo la fracción de población despierta en función de la hora. 

Para su construcción se usa la técnica del \emph{rank-plot} o gráfica por rango. Una vez obtenida una serie de $N$ datos (por ejemplo las $N$ horas de despertar determinadas en los $N$ encuestados de una región) se ordenan los valores de tiempo de menor a mayor obteniéndose una serie $t_1\leq t_2\leq t_3,\dots\leq t_N$. Se representa entonces la posición del valor en la serie (el primer valor es 1, el segundo 2 hasta el último valor que será $N$) frente a su valor temporal. Para comparar series de valores de diferente longitud se divide el valor del rango por $N$ de forma que se obtienen los pares ordenados $(t_1,1/N),(t_2,2/N)\dots (t_N,1)$.

En un punto cualquiera $(t_i,i/N)$ el valor de la coordenada $Y$   representa la fracción de personas que han satisfecho la condición analizada en cualquier instante de tiempo $t\leq t_i$\footnote{Por ejemplo el número de personas que se despiertan antes de la hora $t_i$} respecto del número total de personas. Técnicamente este valor se conoce con el nombre de probabilidad acumulada empírica $P(t<t_i)$.

Ciertos valores de la probabilidad acumulada reciben nombres especiales en Estadística Descriptiva. Así, cuando la probabilidad acumulada empírica alcanza el valor $1/2$  tenemos que la mitad de eventos han ocurrido antes de esa condición temporal y la otra mitad de eventos ocurrirá después de esa condición temporal. Al valor de $t_i$ que hace que la probabilidad empírica valga $1/2$ se le conoce con el nombre de \emph{mediana} o \emph{valor mediano}. La mediana caracterizará las distribuciones temporales de eventos que vamos a analizar.

\section{Resultados}
\label{sec:resultados}

\subsection{En horas civiles}
\label{sec:en-horas-civiles}

\begin{figure*}
  \centering
  \begin{tabular}{cc}
\includegraphics[bb=51 93 280 312]{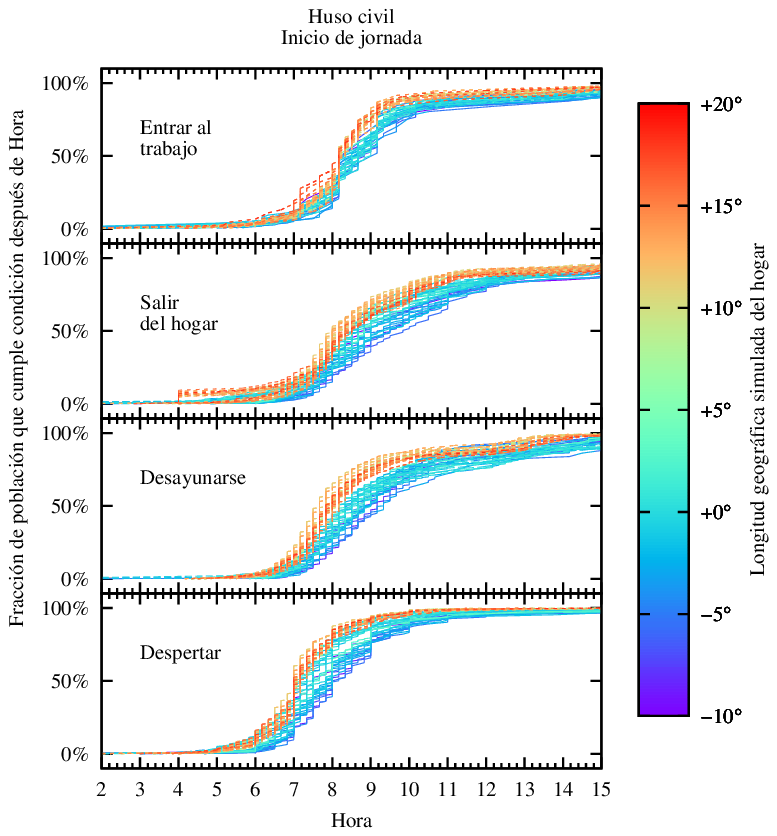}&\includegraphics[bb=75 93 303 312]{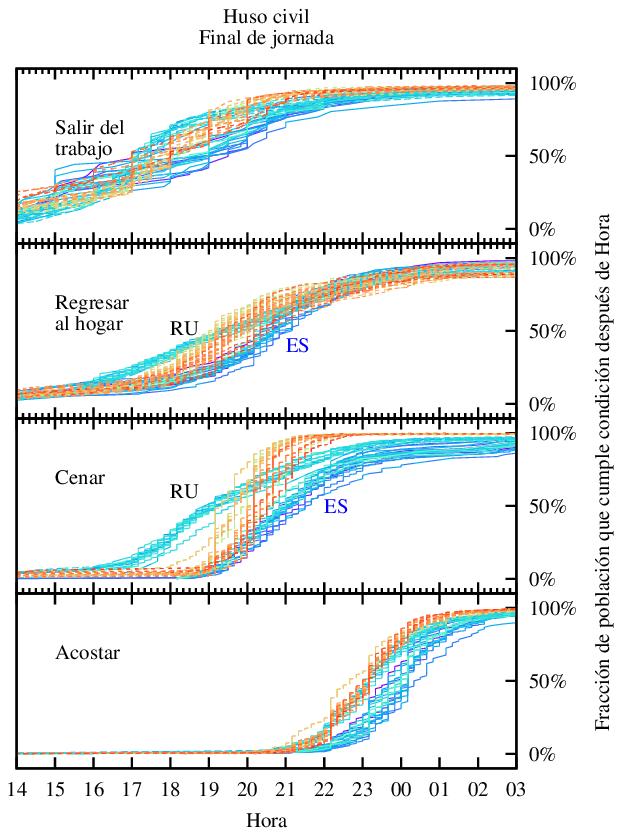}\\
  \end{tabular}
  \caption{Distribución de horas de las actividades y localización analizadas (véase el cuadro~\ref{tab:casos}) y agrupadas por regiones. A la izquierda actividades relacionadas con el inicio de la jornada, a la derecha actividades relacionadas con el fin de la jornada. La hora representada es la civil. El color del trazo señala la longitud geográfica simulada del hogar según la escala cromática mostrada. Canarias se sitúa fuera de la escala cromática y se representa por el color correspondiente a $-10\degree$.}
  \label{fig:InicioCivil}
\end{figure*}

Las gráficas de probabilidad acumulada de los eventos listados en el cuadro~\ref{tab:casos} para las regiones listadas en el cuadro~\ref{tab:numero} con tiempos expresados en forma civil se muestran en la figura~\ref{fig:InicioCivil}.

La longitud geográfica simulada del hogar analizado determina el color con el que se dibuja su dato en la gráfica según una escala cromática que aparece en la figura. Esta escala cromática se extiende desde los $\unit{20}{\degree E}$ hasta los $\unit{10}{\degree W}$. Canarias, que se sitúa más hacia el oeste que el meridiano más occidental de esta escala, se muestra entonces con un color correspondiente a $\unit{10}{\degree W}$.

En estos gráfico, las regiones italianas vienen representadas por un trazo en color cálido, tanto más cálido cuanto más orientales. Las regiones españolas por un trazo azulado tanto más azul cuanto más occidentales; y las regiones británicas quedan en un tono celeste más suave ya que su longitud oeste máxima no alcanza los valores de las regiones españolas (véase el cuadro~\ref{tab:numero}). En los gráficos las regiones españolas y británicas que tienen longitud geográfica parecida están pobremente diferenciadas.

Las características generales de las distribuciones de estas magnitudes son parecidas en las regiones analizadas mostrando en la mayor parte de los casos una forma similar a la probabilidad acumulada de una gaussiana (la función de error). Solo las distribuciones de regreso al hogar y de cenar correspondiente a las regiones británicas son diferentes de las italianas y españolas. Dentro de un mismo país solo las distribuciones de cenar italianas tienen una variabilidad regional interna significativa.

 Es también interesante el hecho de que la distribución de eventos de inicio de jornada es sustancialmente más estrecha que la distribución de eventos de fin de jornada: la gente despierta, se desayuna, sale de casa, entra al trabajo en una franja de pocas horas comparada con la franja en la que la gente cena, regresa al hogar o sale del trabajo. Solo una distribución de fin de jornada ---la de acostar--- es parecida en amplitud a las distribuciones de inicio de jornada. La anchura de las distribuciones se discute en la sección~\ref{sec:amplitud-de-las}.

Las distribuciones de las regiones del Reino Unido relativas al regreso a casa son diferentes de las de Italia y España, mostrando una mayor amplitud.

 Las distribuciones están bien representadas por sus valores medianos: el valor de la variable estadística (en este caso la hora del evento) para el cual la mitad de la población ($50\%$) satisface una determinada condición.  En este cuadro de valores las regiones se listan de oriente a occidente y los eventos se listan aproximadamente en orden cronológico: primero los relacionados con el inicio de la jornada, después los relacionados con el fin de la jornada. Hay que destacar para cada actividad o localización se analiza el grupo de individuos que declaran haberla realizado. Esto es particularmente importante en el caso de "<\utrabajar"> cuyo valor mediano es menor que el valor de la localización "<\ucasa"> porque está última incluye a personas que no trabajan (y que están excluidas de esta estadística) y que, en promedio, permanecen más tiempo en casa.

\begin{figure*}
  \centering
  \begin{tabular}{cc}
    \includegraphics[bb=67 82 242 312]{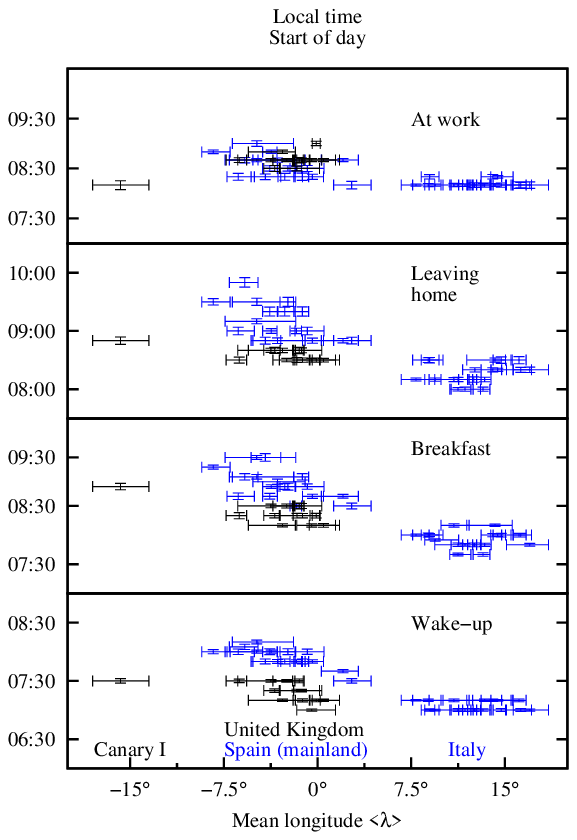}&\includegraphics[bb=73 82 248 312]{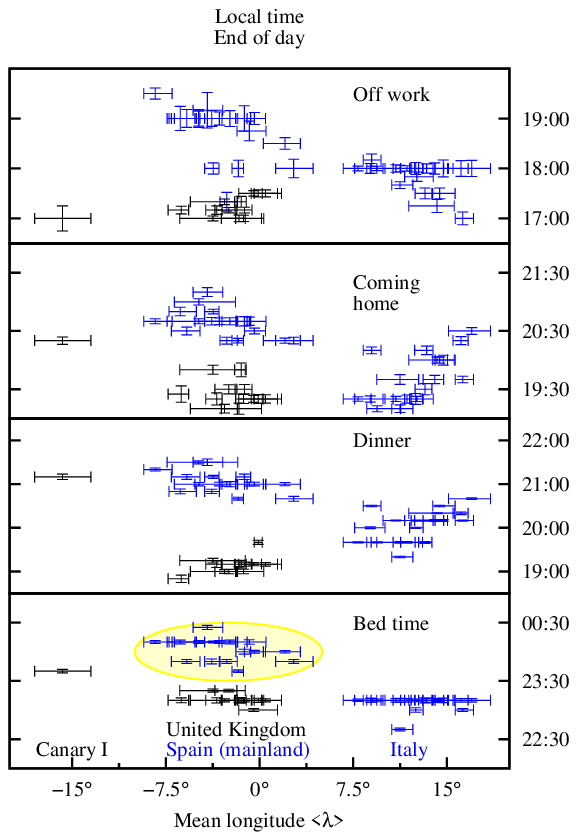}\\   
  \end{tabular}
  \caption{Valores medianos de las distribuciones de horas de inicio (izquierda) y de finalización (derecha) de actividades o localizaciones frente a la longitud geográfica media de la región. Se muestra en negro las regiones con huso WET ---Canarias y las del Reino Unido--- y en azul las que tienen huso CET ---el resto de España e Italia---. Cada región está representada por una línea horizontal trazada desde su longitud más oriental hasta su longitud más occidental y una línea vertical cuya magnitud es una estimación de la incertidumbre del valor mediano analizado. La línea vertical y horizontal se cruzan en cada caso en el valor de la longitud promedio de la región y el valor mediano analizado.}
  \label{fig:Civil}
\end{figure*}

Los valores medianos se presentan gráficamente en la figura~\ref{fig:Civil}. Con carácter general es este tipo de datos el que lleva a asegurar que los horarios españoles son particularmente tardíos\cite{yardley-dinners} y que, por tanto, España vive en un \emph{jet lag} permanente.\cite{vidales-jetlag} Como ejemplo se ha encerrado con un círculo los datos de "<acostar"> de las regiones españolas con huso CET, típicamente más tardía que las de Reino Unido o Italia. Características similares se observan en el resto de estadísticas analizadas en la figura~\ref{fig:Civil} excepto la de inicio de la jornada laboral que presenta unos valores similares para los tres países. En la figura~\ref{fig:Civil} los datos se representen con barras de error: la barra horizontal representa la amplitud geográfica de la región ---corre desde la longitud del municipio regional más occidental al municipio más oriental--- y la barra vertical representa una estimación de la incertidumbre del valor mediano.\footnote{Esta estimación es el cociente entre la desviación absoluta mediana ---véase la sección~\ref{sec:amplitud-de-las}---  dividido por la raíz cuadrada del número de datos que entran en la estadística, que aparecen en el cuadro~\ref{tab:numero}.} Las barras se unen en un punto cuyas coordenadas coincide con la longitud promedio de la región y el valor mediano de la variable.

\subsection{En hora de Londres y en hora solar}
\label{sec:en-hora-de}

El problema de extraer conclusiones a partir de las figuras~\ref{fig:InicioCivil} y~\ref{fig:Civil} ---que los horarios españoles son tardíos, por ejemplo--- es que no hay forma de comparar cabalmente nada en ellas. La dificultad de analizar correctamente estos datos proviene de dos factores diferentes. De una parte las horas civiles no son más que etiquetas, en vez de ser una medida de tiempo con un origen común.\footnote{Por ejemplo las \uhora{07:00WET} y las \uhora{08:00CET} se refieren de forma diferente al mismo instante.} De otra parte hay que tener en cuenta la deriva de la longitud incluso si analizamos horas civiles correspondientes a un mismo huso.

Si comparamos los datos de España con los datos de Italia ---aprovechando que tienen el mismo huso horario salvo en la región canaria---  estamos comparando regiones con casi quince grados de longitud de diferencia, lo que habitualmente se olvida. Si comparamos los datos de España y Reino Unido ---aprovechando que tienen una longitud parecida si, de nuevo, exceptuamos el caso canario--- olvidamos que estas horas son de husos diferentes. Por ejemplo la concurrencia que se observa en la figura relativa a la entrada en el trabajo es solo nominal ---es la misma hora civil mediana--- pero no son eventos simultáneos. En realidad la hora mediana de entrada al trabajo de esas regiones españolas ocurre una hora antes que la hora mediana de entrada al trabajo de las regiones británicas. 

\begin{figure*}[t]
  \centering
  \begin{tabular}{cc}
\includegraphics[bb=73 82 248 312]{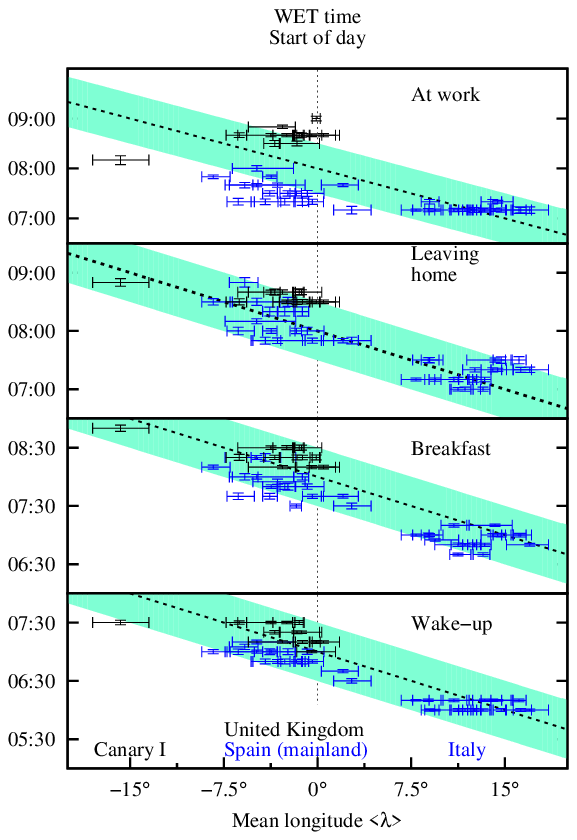}&\includegraphics[bb=67 82 242 312]{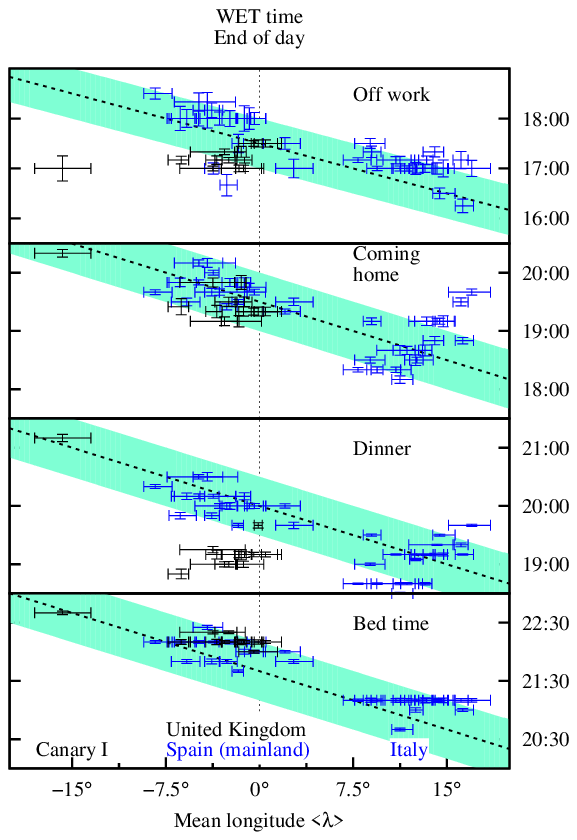}\\   
  \end{tabular}
  \caption{Lo mismo que la figura~\ref{fig:Civil} pero expresando los tiempos tal y como los mediría un observador fijo situado en Londres. Se muestra en negro las regiones con huso WET ---Canarias y las del Reino Unido--- y en azul las que tienen huso CET ---el resto de España e Italia---. Los trazos discontinuos muestran líneas de horas solares constantes. La diferencia entre una línea discontinua y la siguiente es media hora solar. La hora solar crece al desplazarse verticalmente hacia arriba de una a otra.}
  \label{fig:London}
\end{figure*}

Una forma más coherente de analizar estos datos es representar los valores medianos en una baste de tiempo común para todos los datos: por ejemplo la de Londres. Es decir, restar una hora a los valores de las regiones cuyo huso civil sea CET (UTC+01) y que se representan en azul en la figura~\ref{fig:Civil}. Cuando se hace esta operación se obtiene el resultado que muestra la figura~\ref{fig:London} en la que las horas etiquetadas en los ejes verticales son verdaderamente cronológicas. Debe tenerse un especial cuidado a la hora de interpretar esta gráfica: no muestra los horarios que se observarían en España si, hipotéticamente, el gobierno español decidiera retrasar una hora el huso civil vigente; muestra los horarios actuales observados desde un sujeto situado en Londres.

La única diferencia entre la figura~\ref{fig:Civil} y la figura~\ref{fig:London} es que los valores representados en azul (correspondientes a las regiones de Italia y España, salvo Canarias) se desplazan una hora hacia abajo. Entonces todos los datos de valores medianos representados muestran claramente una dependencia lineal con la longitud geográfica en contraste con lo que ocurre en la figura~\ref{fig:Civil}.

Aquí es evidente que las regiones más orientales realizan sus actividades cronológicamente antes que las regiones más occidentales. Esta tendencia podría representarse por el valor de una pendiente obtenida por un método de regresión lineal. Sin embargo en este caso es mucho más instructivo analizar el fenómeno subyacente a esta dependencia, que es la evolución del tránsito solar. Esta evolución se realiza a un ritmo constante $\nu$ de cuatro minutos de adelanto por grado de longitud geográfica hacia el este.\footnote{Esta constante es la expresión en minutos por grado sexagesimal del tiempo $\tau$ que tarda la Tierra en dar una revolución alrededor de sí misma; siendo $1/\nu=\tau/2\pi$}  Por ello en la figura se muestra, con líneas discontinuas, rectas cuya pendiente es justamente $-\nu$. Estas rectas representan horas solares constantes.\footnote{Recordemos que la "<hora solar"> expresa la distancia temporal de un evento respecto del tránsito del Sol por el meridiano antipodal al meridiano del observador; es decir, respecto de la medianoche local. Como la medianoche y el mediodía se diferencian exactamente en $\tau/2$, la hora solar marca también la distancia al mediodía solar local.} La banda de color que muestran las gráficas representa una banda solar de una hora de amplitud. Entonces lo que se observa en la figura~\ref{fig:London} es que los valores medianos analizados se distribuyen alrededor de una banda de una hora solar con independencia de que cronológicamente se extiendan en más tiempo.\footnote{Debe destacarse que haber dibujado estas rectas de pendiente $-\nu$ en la figura~\ref{fig:Civil} no tendría sentido ya que las horas civiles no se corresponden con una medida de tiempo tomada por un observador único y no establecen una cronología de eventos. Por eso, los datos de esa figura no muestran ninguna tendencia lineal.}

\begin{figure*}[t]
  \centering
  \begin{tabular}{cc}
        \includegraphics[bb=67 82 242 312]{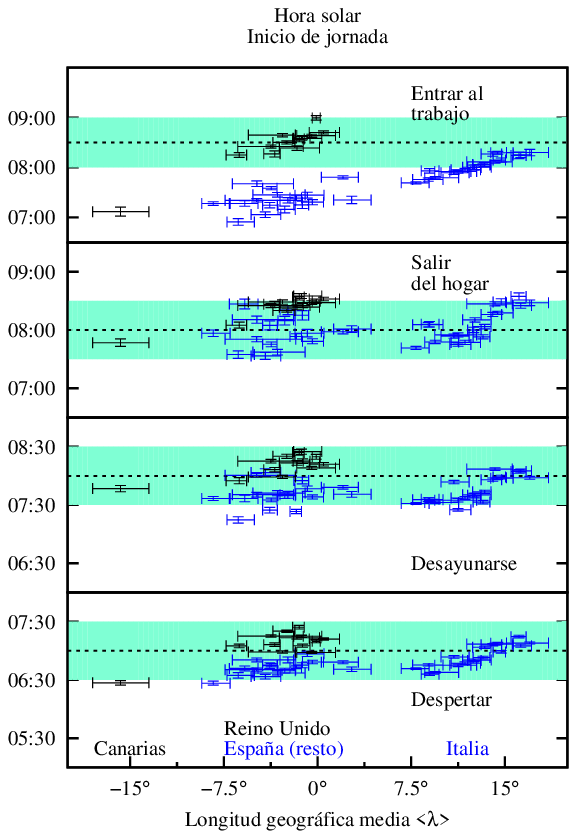}&\includegraphics[bb=73 82 248 312]{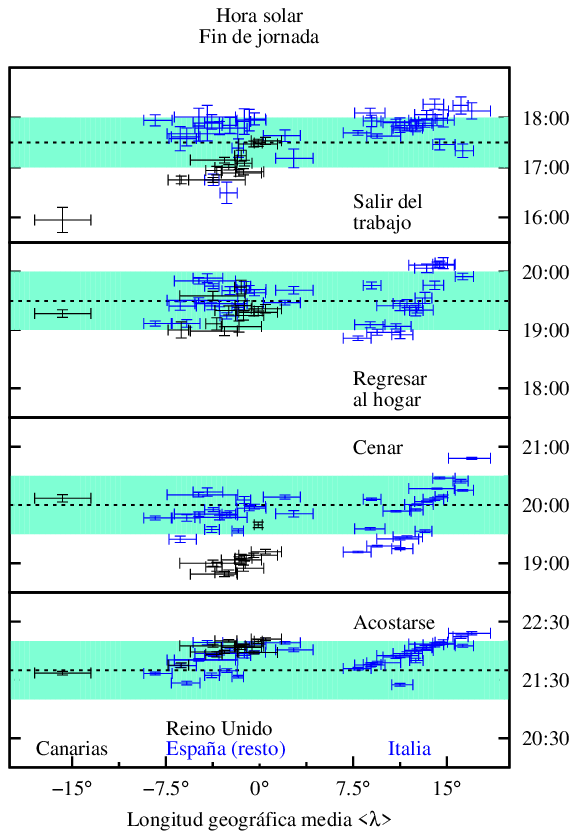}\\
  \end{tabular}
  \caption{Lo mismo que las figuras~\ref{fig:Civil} y~\ref{fig:London} pero con los tiempos expresados en horas solares, es decir corregidas por la longitud geográfica de la región. Se muestra en negro las regiones con huso WET ---Canarias y las del Reino Unido--- y en azul las que tienen huso CET ---el resto de España e Italia---. Los trazos discontinuos muestras líneas de horas solares constantes.}
  \label{fig:solar}
\end{figure*}

El valor mediano de la hora solar a la que ocurren los eventos puede calcularse sin más que aplicar la ecuación~(\ref{eq:1}) al valor mediano representado en la figura~\ref{fig:London}. La longitud geográfica se tomará como el valor promedio de la región a la que se refiera el valor mediano. El resultado de esta operación se muestra en la figura~\ref{fig:solar}.

La figura~\ref{fig:London} presenta una cronología de eventos ya que la medida de tiempo es externa con un reloj fijo. La figura~\ref{fig:solar} no, ya que la medida de tiempo es intrínseca: se representan los valores medianos en hora solar local de cada región. En compensación, la figura~\ref{fig:solar} permite trazar la posición del Sol respecto de los horarios observados.

La figura~\ref{fig:solar} es en realidad un promedio anual del muestreo de actividades sobre las regiones analizadas ya que no dividimos las respuestas por meses o por trimestres. El orto y el ocaso solar dependen del día del año pero su valor promedio anual es, aproximadamente, el correspondiente a los equinoccios. Es decir las \uhora{06:00} para el orto y las \uhora{18:00} para el ocaso. El empleo del horario de verano retrasa estos valores medios nominales media hora aproximadamente, ya que el horario de verano está vigente medio año\footnote{En realidad el horario de verano está vigente $7/12$ partes del año, y el de invierno $5/12$}. Así el orto promedio anual es cercano a las \uhora{06:30} y el ocaso promedio anual a las \uhora{18:30}.

La figura~\ref{fig:solar} muestra por tanto que todas la regiones analizadas se despiertan, en promedio, poco después de amanecer. También que en todas las regiones se regresa al hogar poco después de la puesta del sol. Ambas cosas parecen  bastante naturales.

Otro aspecto interesante de la figura~\ref{fig:solar} es la posición relativa de los horarios españoles. Si nos fijamos en las actividades de inicio de jornada se observa que si algo caracteriza a estos horarios no es el hecho de que sean tardíos sino, todo lo contrario, que son relativamente más tempranos que los observados en Italia o el Reino Unido ya que se sitúan sobre la recta de hora solar constante más temprana; este asunto se abordará posteriormente en la sección~\ref{sec:influencia-de-la}.

Si centramos nuestra atención en los datos de finalización de la jornada hay una sincronización con los datos italianos y una  mayor variabilidad con los horarios británicos. Estos en general son más tempranos cuando analizamos la salida del trabajo o la cena, pero son coincidentes si nos fijamos en el regreso al hogar o en la hora de acostarse.

En la figura~\ref{fig:solar} se observa que el horario de inicio de la actividad laboral en las regiones españolas es singularmente más temprano que en Italia y Reino Unido. Esto se analizará en la sección~\ref{sec:influencia-de-la}.

\subsection{Intervalos de tiempo}
\label{sec:tiempo-acumulado-de}

Los intervalos de tiempo ---la distancia temporal entre una marca inicial y una marca final--- no dependen del horario en el que se mida una actividad. Por tanto no cabe diferenciar en este caso entre tres relojes diferentes. Por tanto el tratamiento de este tipo de estadísticas es diferente de las estadísticas horarias anteriores. Por ejemplo, y en principio, no cabría esperar ahora efectos relacionados con la longitud. En este estudio se analizan dos intervalos de tiempo: la amplitud de las distribuciones horarias y el tiempo acumulado de actividad.

\subsubsection{Amplitud o variabilidad de las distribuciones}
\label{sec:amplitud-de-las}

La amplitud de las distribuciones horarias que aparecen en la figura~\ref{fig:InicioCivil} es el intervalo de tiempo aproximado en el que la distribución pasa de tener un valor pequeño ---poca población ha cumplido una condición--- a un valor alto ---casi toda la población ha cumplido una condición. Para estimar esta anchura, amplitud o variabilidad de la estadística se computa primero la cantidad
\begin{equation}
  \label{eq:5}
  T_i=\mathrm{abs}(t_i-\langle t\rangle),
\end{equation}
donde $\mathrm{abs}(x)$ es la función valor absoluto de $x$ y $\langle t\rangle$ es el valor mediano de la actividad. La estadística $T_i$ guarda la diferencia absoluta entre los valores de tiempo observado y el valor mediano del conjunto. Como la ecuación~(\ref{eq:5}) computa una diferencia de tiempo los valores de $T_i$ no dependen de ninguno de los factores asociados a la hora y el huso horario.

Valores de $t_i$ cercanos al valor mediano $\langle t\rangle$ dan valores de $T_i$ cercanos a cero. Valores de $t_i$ alejados del valor mediano por cualquier banda ---anterior o posterior--- dan valores grandes de $T_i$. El valor mediano de la estadística $T_i$ se denomina \emph{desviación absoluta mediana} y es una medida razonable de la anchura, amplitud o variabilidad de la distribución. Si la distribución es aproximadamente simétrica, el 75\% de los valores observados entrará en una banda cuyo valor central es el valor mediano, y cuya amplitud es la desviación absoluta mediana.

\begin{figure*}
  \centering
  \begin{tabular}{cc}
        \includegraphics[bb=67 82 242 312]{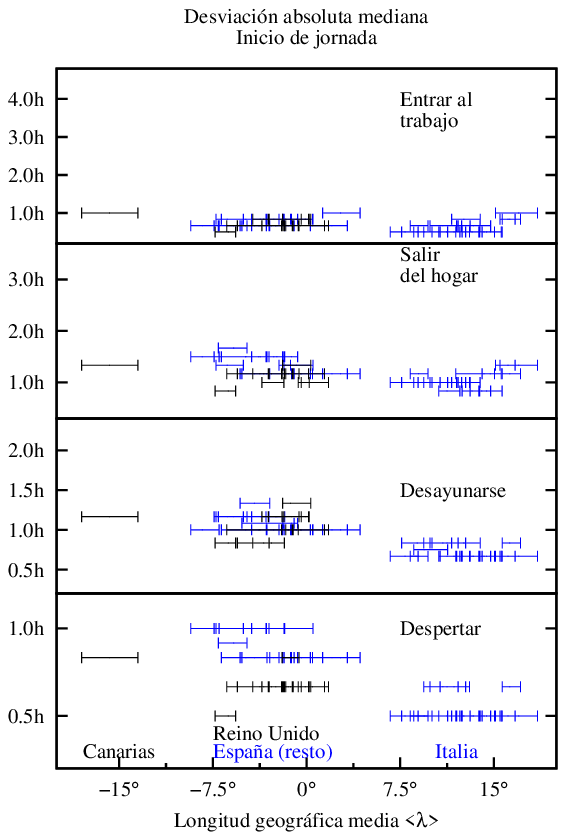}&\includegraphics[bb=73 82 248 312]{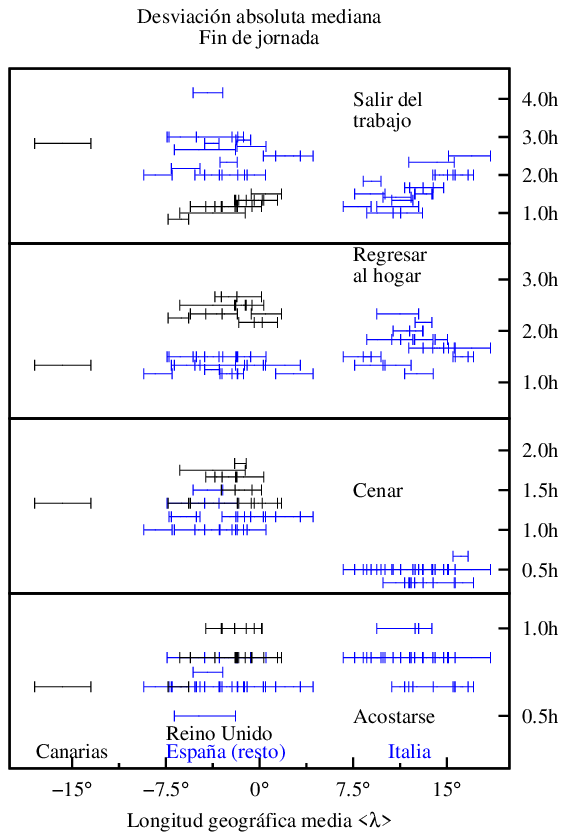}\\
  \end{tabular}
  \caption{Anchura de las distribuciones observadas en la figura~\ref{fig:InicioCivil} determinadas por el valor mediano de Ec.~(\ref{eq:5}). Obsérvese como la anchura de las distribuciones de inicio de jornada es en general menor que la anchura de las distribuciones de fin de jornada. Obsérvese también la mayor variabilidad de la estadística de salida del trabajo.}
  \label{fig:dispersion}
\end{figure*}

La figura~\ref{fig:dispersion} muestra la desviación absoluta mediana de las muestras analizadas. En contraste con las estadísticas horarias, la figura~\ref{fig:dispersion} no muestra dependencia con la longitud geográfica ---compárese con la figura~\ref{fig:London}. Sí hay un cierto grado de variabilidad de un país a otro y de una actividad a otra. Las distribuciones de salida de trabajo y de regreso al hogar tienen valores de la desviación absoluta mediana especialmente elevados. Esto sugiere que el inicio de jornada ---despertar y empezar a trabajar--- es un fenómeno más repentino, mientras que el final de jornada es más escalonado y progresivo.

\subsubsection{Tiempo acumulado de una actividad o localizaci{\`o}n}
\label{sec:tiempo-acumulado-de-1}

El otro intervalo de tiempo que puede analizarse es el recuento del tiempo acumulado de una actividad. Es decir cuánto tiempo dormimos o cuánto tiempo trabajamos. Este tiempo acumulado de actividad es la suma total de todos los intervalos de tiempos en los que realizamos la actividad. Por tanto no dependen de consideraciones relativas al tipo de hora en el que se realiza la medida. 

\begin{figure*}
  \centering
  \begin{tabular}{cc}
        \includegraphics[bb=67 82 242 312]{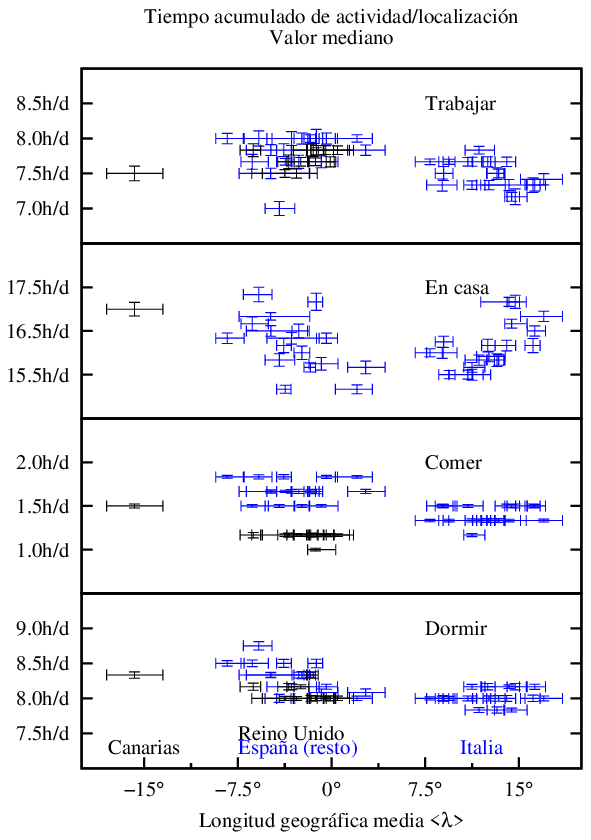}&\includegraphics[bb=73 82 248 312]{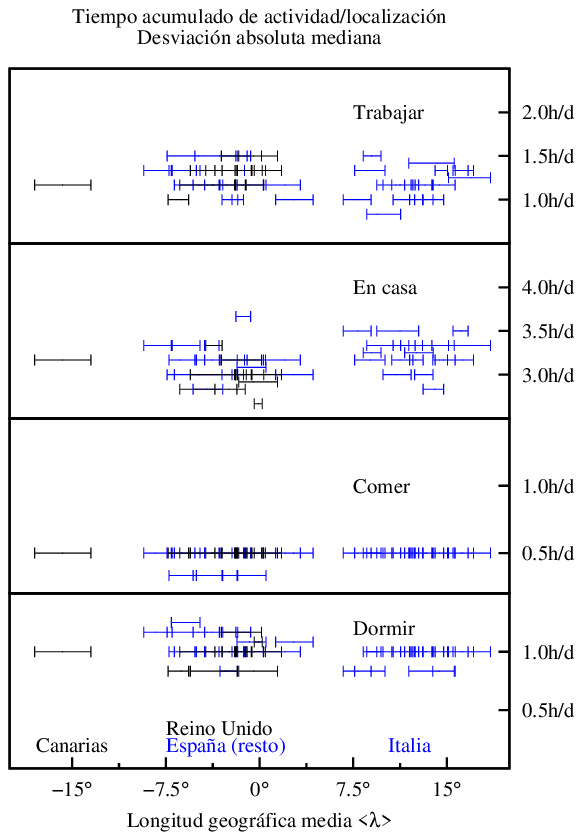}\\
  \end{tabular}
  \caption{A la izquierda valores medianos de las distribuciones de tiempo acumulado realizando una actividad o localización. A la derecha anchura de las distribuciones. Como es de esperar para la medida de intervalos de tiempo no hay correlación de los valores representados con la longitud geográfica media.}
  \label{fig:acumulado}
\end{figure*}

La figura~\ref{fig:acumulado} muestra los valores medianos (a la izquierda) y la desviación absoluta mediana (a la derecha) del tiempo total empleado en las actividades listadas en el cuadro~\ref{tab:casos}.

Los valores medianos son prácticamente coincidentes en las regiones de los tres países analizados.

Debe observarse que el tiempo acumulado, como su propio nombre indica, no es sensible a cómo se distribuye una actividad a los largo del día. Es posible que los encuestados repartan en varios turnos sus actividades de dormir o trabajar.

\section{Discusi{\'o}n}
\label{sec:discusion}
 
¿Cuál es la interpretación de los resultados mostrados en la sección~\ref{sec:resultados}?

Los resultados principales están contendidos en la figura~\ref{fig:London}, referido a los horarios de las actividades, y en la figura~\ref{fig:acumulado}, referido al tiempo acumulado de las actividades.

La figura~\ref{fig:London} muestra que los horarios de las actividades de las encuestas analizadas varían principalmente por el efecto natural de la revolución aparente del Sol alrededor de la Tierra. En la figura puede observarse como todas las estadísticas analizadas, con las excepciones de la hora de entrada al trabajo y la hora de la cena, se sitúan en una banda de una hora solar.

Que los valores medianos observados en las tres encuestas se incluyan en una banda de una hora de anchura solo puede calificarse de normal y aceptable teniendo en cuenta que trabajamos con un sistema de husos horarios cuya amplitud es una hora y teniendo en cuenta que la decisiones que toman las personas respecto a los horarios suelen estar discretizadas en valores enteros o medios de horas.

Por tanto, la figura~\ref{fig:London} muestra que los horarios españoles \emph{ya} están adaptados al huso vigente, de forma que las regiones españolas realizan sus actividades a unas horas que si bien son nominalmente anómalas y tardías, son también totalmente compatibles con la posición del Sol y comparables con los horarios a las que se realizan en Italia o Gran Bretaña. Es decir, aunque España observe el huso horario CET ---desplazado quince grados al este de su posición geográfica--- realiza sus actividades de acuerdo con su meridiano, es decir el huso horario WET. O lo que es lo mismo España no vive bajo un \emph{jet lag} permanente.

La figura~\ref{fig:acumulado} muestra que los tiempos acumulados de las tres encuestas son totalmente comparables unos con otros e independientes de la longitud. La diferencia de valores medianos de las estadísticas de Italia, España y Reino Unido es menor que las desviaciones absolutas medianas de estas mismas estadísticas.

\subsection{Influencia de la latitud: por qu{\'e} los espa{\~n}oles madrugan}
\label{sec:influencia-de-la}

Una de las ventajas de usar la hora solar es que en ella se contiene toda la dependencia del problema respecto de la longitud. Es decir, del movimiento aparente del Sol en el cielo en el transcurso de un día. 

Uno de los aspectos más llamativos de la figura~\ref{fig:solar} ---o de la figura~\ref{fig:London}--- es que el valor mediano de los horarios de entrada al trabajo de los españoles es significativamente más temprano que el de italianos o británicos. 

\begin{figure*}[t]
  \centering
  \begin{tabular}{cc}
\includegraphics[bb=73 82 248 312]{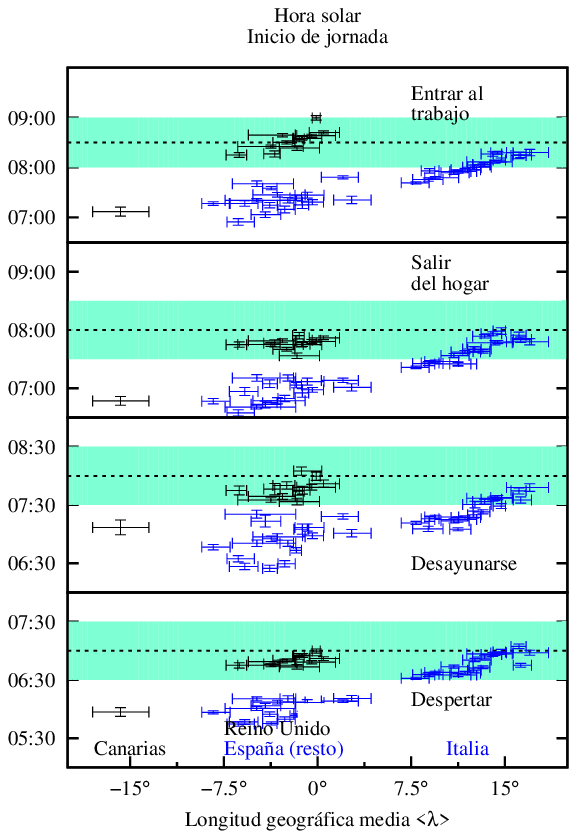}&\includegraphics[bb=67 82 242 312]{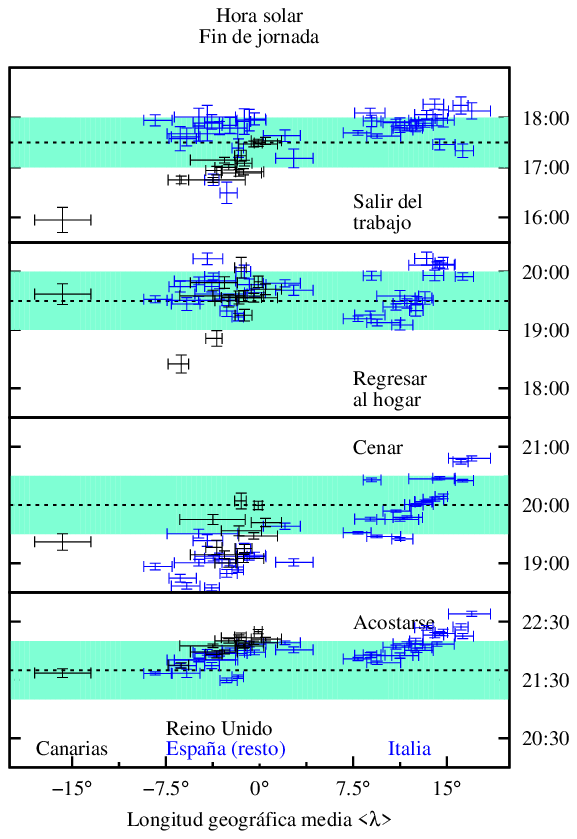}\\   
  \end{tabular}
  \caption{Valores medianos de las estadísticas de inicio y fin de jornada computadas para el conjunto de personas que trabajan ---véase el cuadro~\ref{tab:numero}. Obsérvese cómo los valores de las regiones españolas de inicio de jornada son sustancialmente más tempranos que los de las regiones italianas y británicas.}
  \label{fig:LondonTrabaja}
\end{figure*}

En estas figuras la población de la muestra es el campo disponible en el conjunto de datos y que se lista en el cuadro~\ref{tab:numero}. Si se restringe el análisis al conjunto de personas que trabajan se obtiene el resultado que muestra la figura~\ref{fig:LondonTrabaja}. Aquí, la desviación de los valores españoles en las estadísticas de inicio de jornada es significativamente diferente. Sin embargo los valores de fin de jornada son coincidentes.

La cuestión es ¿por qué las regiones españolas tienen horarios de inicio más tempranos?  La explicación más recurrente es que esto es un efecto  del cambio horario habido en el año 1940. En general un cambio como este se hace para "<madrugar más"> y realizar antes las actividades. En la sección~\ref{sec:el-trasf-hist} se argumentará cómo esto es solo válido inicialmente y que se podría esperar que con el paso del tiempo el efecto madrugador se amortiguara.

\emph{Stricto sensu} culpar al cambio de hora del efecto observado en la figura~\ref{fig:LondonTrabaja} puede ser una falacia \emph{post hoc ergo propter hoc} si lo único que se conociera fuera la secuencia: primero, cambio de huso horario en el año 1940; segundo observamos horarios anómalos en el rango 1990-2000. Y concluimos que lo primero causa lo segundo. 

Una de las ventajas de corregir los valores horarios medidos por la longitud es que ahora tenemos una variable ---los valores medianos expresados en hora solar--- expresada en una base común para todas las regiones observadas. La variación de estos valores medianos ya no puede deberse al factor longitud sino que debe relacionarse con otros factores. 

\subsubsection{{¿}C{ó}mo influye la latitud en el movimiento aparente del Sol?}
\label{sec:como-influye-la}

La longitud geográfica influye en el movimiento diario del Sol y condiciona la secuencia cronológica de eventos. Como el Sol se mueve de oriente a occidente es mediodía antes en las regiones orientales que en las occidentales. Y, en general, todo la secuencia de eventos queda gravemente condicionada por este hecho.

La latitud no influye en este ciclo diario pero sí en el ciclo anual del movimiento de la Tierra alrededor del Sol condicionado por la inclinación del eje de rotación de la Tierra respecto del plano de traslación. Esta inclinación, llamada oblicuidad, tiene actualmente un valor de $\epsilon=\unit{23.5}{\degree}$. En este ciclo anual la latitud del punto subsolar ---que es el punto sobre el que inciden perpendicularmente los rayos del Sol--- más comúnmente llamada \emph{declinación solar} pasa de valer $\delta_s=\unit{0}{\degree}$ en los equinoccios ---el Sol está sobre el Ecuador--- a valer $\delta_s=\pm\epsilon$ en los solsticios ---el valor positivo para el solsticio boreal y el valor negativo para el solsticio austral. En una descripción simple $\delta_s$ es una magnitud que varía cíclicamente a lo largo del año y es función del día.

El valor de la latitud $\phi$ y la declinación solar $\delta_s$ condicionan la insolación; que fundamentalmente es función de $\cos(\phi-\delta_s)$. Influye también en el ángulo paraláctico que es el ángulo que forma la trayectoria del Sol con el horizonte. E influye también en la hora del amanecer.

Los dos primeras influencias son monótonas en el año y, en general, cuanto mayor es la latitud menor es la insolación y menor es el ángulo paraláctico\footnote{El menor ángulo paraláctico hace que los amaneceres y atardeceres sean más lentos conforme aumenta la latitud: desde el momento del amanecer el Sol gana altura más lentamente cuando mayor sea la latitud.} a lo largo de todo el año. 

La hora del amanecer tiene un comportamiento periódico. En verano es tanto más temprana cuando mayor es la latitud; en invierno lo contrario. 

El promedio anual de la hora del amanecer es aproximadamente independiente de la latitud, siempre que no se consideren latitudes polares, y coincide con el cuarto de día; es decir las $\uhora{06:00}$ hora del meridiano local de observación. El cambio de hora en verano hace que este valor esté desplazado hasta las $\uhora{06:30}$ si computamos la hora solar respecto del meridiano local en invierno y respecto del meridiano local más $\unit{15}{\degree}$ en verano. 

Sin embargo, es la hora del amanecer quien más condiciona los horarios laborales.

En cualquier instante el ángulo $z$ que forma el centro del disco solar con el horizonte está determinado por las coordinadas geográficas del observador $\lambda,\phi$ y por las coordenadas geográficas del punto subsolar $\lambda_s,\phi_s$ a través de la relación:\cite{boeker-grondelle-2011,wiki-sunriseequation}
\begin{equation}
  \label{eq:4}
  \sin z=(\cos(\lambda-\lambda_s)+\tan\phi\tan\phi_s)\times\cos\phi\cos\phi_s.
\end{equation}
 El punto subsolar es el punto de la Tierra sobre el cual los rayos del sol inciden perpendicularmente; es decir el punto cuya altura angular es $z=\unit{90}{\degree}$. Las coordenadas del punto subsolar define la longitud geográfica $\lambda_s$ y latitud geográfica $\phi_s$ del Sol. A esta latitud se le denomina normalmente \emph{declinación solar} y se simboliza por $\delta_s$. 

La longitud solar $\lambda_s$ va variando a lo largo del día según el movimiento de aparente del sol. En el lapso de un día $\delta_s$ prácticamente no cambia, pero a lo largo del año pasa del valor extremo $\delta_s=+\epsilon$ en el solsticio boreal, al valor extremo $\delta_s=-\epsilon$ en el solsticio austral, pasando dos veces por el valor medio $\delta_s=\unit{0}{\degree}$. La constante $\epsilon$ es la inclinación del eje de rotación de la Tierra respecto del plano de traslación y, actualmente, vale $\unit{23.5}{\degree}$. El movimiento anual de la declinación solar puede representarse por:
\begin{equation}
  \label{eq:9}
  \sin\delta_s=\sin\epsilon\sin\left(\frac{2\pi}{T}(d-d_0)\right),
\end{equation}
donde $d$ es el número de días que ha transcurrido desde el principio del año; $T=\unit{365}{\uday}$ es el periodo del movimiento de traslación de la Tierra alrededor del Sol ---la duración del año--- y $d_0=\unit{91}{\uday}$ es el número de días que transcurren desde el principio del año hasta el día del equinoccio de marzo.\footnote{La expresión~(\ref{eq:9}) es aproximada. El movimiento de traslación de la Tierra es más complejo ya que, entre otras cosas, la Tierra se acelera o frena según su posición dentro de la órbita elíptica que describe. En la ecuación~(\ref{eq:9}) se supone que el movimiento es circular y uniforme.}

En la ecuación~(\ref{eq:4}) $z$ alcanza diariamente valores extremos cuando $|\cos(\lambda_s-\lambda)|=1$. Es decir en el mediodía local ($\lambda_s-\lambda=0$) y en la medianoche local ($\lambda_s-\lambda=\unit{180}{\degree}$). Y alcanza anualmente valores extremos cuando $\phi_s=\pm\epsilon$; es decir en los solsticios boreal ($\delta_s=\epsilon$) y austral ($\delta_s=-\epsilon$). 

El ángulo $\lambda-\lambda_s$ es la distancia angular que separa el meridiano del punto subsolar del meridiano del observador. Si consideramos la razón $\nu=\tau/2\pi$ ---los cuatro minutos por grado ya antes utilizados--- se tiene que $\nu(\lambda-\lambda_s)$ convierte la distancia angular en distancia temporal respecto del mediodía local:
\begin{equation}
  \label{eq:12}
  \nu(\lambda-\lambda_s)=\tau/2\pm t
\end{equation}
donde $t$ es la hora solares. Entonces de la ecuación~(\ref{eq:4}) se obtiene la ecuación de evolución solar como:
\begin{equation}
  \label{eq:6}
  t=\frac{\tau}{2}\pm\nu\cos^{-1}\left(\tan\phi\tan\delta_s-\frac{\sin z}{\cos\phi\cos\delta_s}\right),
\end{equation}
que relaciona la hora solar del día, el día del año ---a través de la ecuación~(\ref{eq:9})--- y la latitud del observador con la altura del Sol sobre el horizonte en ese instante. El signo $\pm$ se refiere a momentos \emph{post meridiem} ($+$) o \emph{ante meridiem} ($-$) ya que el mediodía $t=\tau/2$ divide el día en dos mitades iguales ---como su nombre indica.

En la ecuación~(\ref{eq:6}) los valores extremos de $z$ siguen ocurriendo para $\delta_s=\pm\epsilon$, es decir en los solsticios.

\subsubsection{Aqu{í} amanece muy temprano}
\label{sec:aqui-amanece-muy}

Cuando la altura angular del Sol sobre el horizonte se anula la ecuación~(\ref{eq:4}) se simplifica y se obtiene la condición:
\begin{equation}
  \label{eq:7}
  \cos(\lambda-\lambda_s)=-\tan\phi\tan\delta_s
\end{equation}
La condición $z=0$ determina, aproximadamente, el orto y el ocaso solar. Si se considera el tamaño angular del Sol ---un disco de medio grado sexagesimal--- y la influencia de la refracción de los rayos del Sol en la atmósfera terrestre se obtiene que el momento en el que el limbo superior del Sol aparece o desaparece por el horizonte está determinado por la condición $z_0=-\unit{0.83}{\degree}$ ---cuando el centro del Sol está ochenta y tres décimas de grado por debajo del horizonte,--- que de momento despreciaremos.

La doble condición de orto y ocaso aparece en la solución de la ecuación~(\ref{eq:7}) por la simetría de la función coseno ---$\cos(x)=\cos(-x)$--- de forma que la ecuación~(\ref{eq:7}) se satisface para dos valores de $\lambda_s$ equidistantes de $\lambda$ tales que:
\begin{equation}
  \label{eq:11}
  |\lambda-\lambda_s|=\cos^{-1}(-\tan\phi\tan\delta_s)
\end{equation}
El valor más oriental es el amanecer, el más occidental el anochecer. 

La ecuación~(\ref{eq:11}) es extremal también cuando $\delta_s=\pm\epsilon$ obteniéndose de ella la relación extremal entre latitud y longitud para el orto y el ocaso del Sol:
\begin{equation}
  \label{eq:8}
  |\lambda-\lambda_s|=\cos^{-1}(\mp\tan\phi\tan\epsilon)
\end{equation}
que determina la relación entre la latitud del observador y la longitud del observador en el amanecer del solsticio austral.

\begin{figure*}
  \centering
 \includegraphics[bb=50 50 394 268]{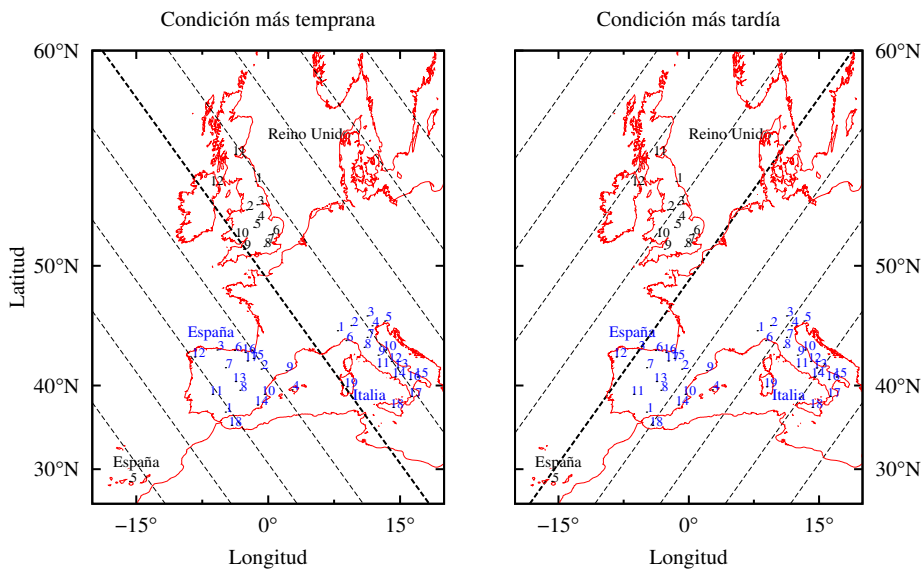}
  \caption{Mapa de Europa occidental con las líneas de orto y ocaso solar más tardías y más tempranas. La latitud está distorsionada de forma que estas líneas sean rectas. A la derecha el amanecer ---verano--- y el anochecer ---invierno--- más tempranos. A la izquierda, el amanecer ---invierno--- y el anochecer ---verano--- más tardío. La familia de rectas se diferencian en el valor de $\lambda_s$. Entre una línea y la siguiente hay $\unit{7.5}{\degree}$ de diferencia en el valor de $\lambda_s$ que equivalen exactamente a media hora de tiempo.}
  \label{fig:extremal}
\end{figure*}
La figura~\ref{fig:extremal} muestra el mapa de Europa occidental con el eje de la latitud distorsionado por la función $\cos^{-1}(\tan\phi\tan\epsilon)$. De esta forma las líneas extremas de salida y puesta del Sol aparecen como rectas. En la gráfica de la izquierda se muestra la condición más temprana de puesta y salida de Sol. A la derecha la condición más tardía. Debido a la simetría de la ecuación~(\ref{eq:8}) la línea de amanecer más tardío ---que ocurre en invierno--- coincide con la línea del anochecer más tardío ---que se da en verano. La única diferencia es que para el amanecer las líneas dejan el día por la derecha y la noche por la izquierda; mientras que para el anochecer es al revés. 

La familia de rectas representada en la figura se diferencia en el valor de $\lambda_s$ que van incrementándose un valor constante $\unit{7.5}{\degree}$ de recta a recta. Esto supone una diferencia temporal de media hora entre el amanecer que representa una recta y el que representa la siguiente.

En la gráfica de la derecha la línea más gruesa señala el amanecer del solsticio de invierno cuando $\lambda_s=\unit{60}{\degree}$. Es decir, cuando el Sol incide perpendicularmente sobre las coordenadas $\lambda_s=\unit{60}{\degree E}$ y $\delta_s=\unit{23.5}{\degree S}$ ---en algún lugar del océano Índico, al sureste de las islas de Reunión y de Mauricio--- amanece sobre la línea gruesa de la gráfica de la derecha que pasa por Asturias (3), por Canarias (5) y por el borde costero de los Países Bajos. Este amanecer ocurre a las $\uhora{08:00UTC}$ porque:
\begin{equation}
  \label{eq:10}
  \frac{\unit{24}{\hour}}{2\pi}\times\frac{2\pi}{\unit{360}{\degree}}\times|\unit{60}{\degree E}-\unit{180}{\degree E}|=\unit{8}{\hour}
\end{equation}
es decir, han pasado ocho horas desde que el Sol atravesó el meridiano antipodal a Greenwich. Alternativamente, la misma línea de la gráfica derecha muestra el anochecer de verano cuando $\lambda_s=\unit{120}{\degree W}$\footnote{El valor de $\lambda_s$ en invierno y el valor de $\lambda_s$ en verano se diferencian en $\unit{180}{\degree}$.} y $\delta_s=\unit{23.5}{\degree N}$; un punto del oceáno Pacífico al oeste del cabo San Lucas en la Baja California. La hora de este anochecer es, lógicamente, las \uhora{20:00}{UTC}.\footnote{La línea más gruesa de la gráfica de la izquierda muestra la situación simétrica. El amanecer más temprano cuando el Sol está a $\unit{120}{\degree E},\unit{23.5}{\degree N}$ en el borde occidental de la isla de Taiwan a las $\uhora{04:00UTC}$, y el anochecer más temprano cuando el Sol está a $\unit{60}{\degree W},\unit{23.5}{\degree S}$ al oeste de Paraguay y son las $\uhora{16:00UTC}$.}

La cuestión importante a destacar en el gráfico~\ref{fig:latitud} (izquierda) es que aunque esté amaneciendo a la vez en Asturias (3) que en Canarias (5) el tiempo que queda para que se produzca el mediodía en ambos lugares es diferente. Efectivamente el Sol está en $\unit{60}{\degree E}$, Asturias tiene una longitud media de $\unit{5.82}{\degree W}$ y Canarias $\unit{15.8}{\degree W}$. Entonces falta $\unit{4.4}{\hour}$ para el mediodía en Asturias y $\unit{5}{\hour}$ en Canarias. Es decir, la hora solar de ese instante es diferente en una comunidad y en otra.

Por tanto, la hora solar da información sobre la distancia al mediodía o a la medianoche que tiene un evento pero no da información respecto de la distancia al orto o al ocaso solar. Tampoco da información sobre la altura $z$ que tiene el Sol sobre el horizonte. Mismos valores de horas solares representan situaciones muy diferentes de la posición del Sol sobre el cielo. Y al revés, valores distintos de la hora solar pueden referirse a situaciones similares de la posición del Sol en el cielo.

\subsubsection{Los valores medianos de entrada al trabajo y la luz del Sol}
\label{sec:los-valores-medianos}

\begin{figure*}[t]
  \centering
  \includegraphics[scale=0.9]{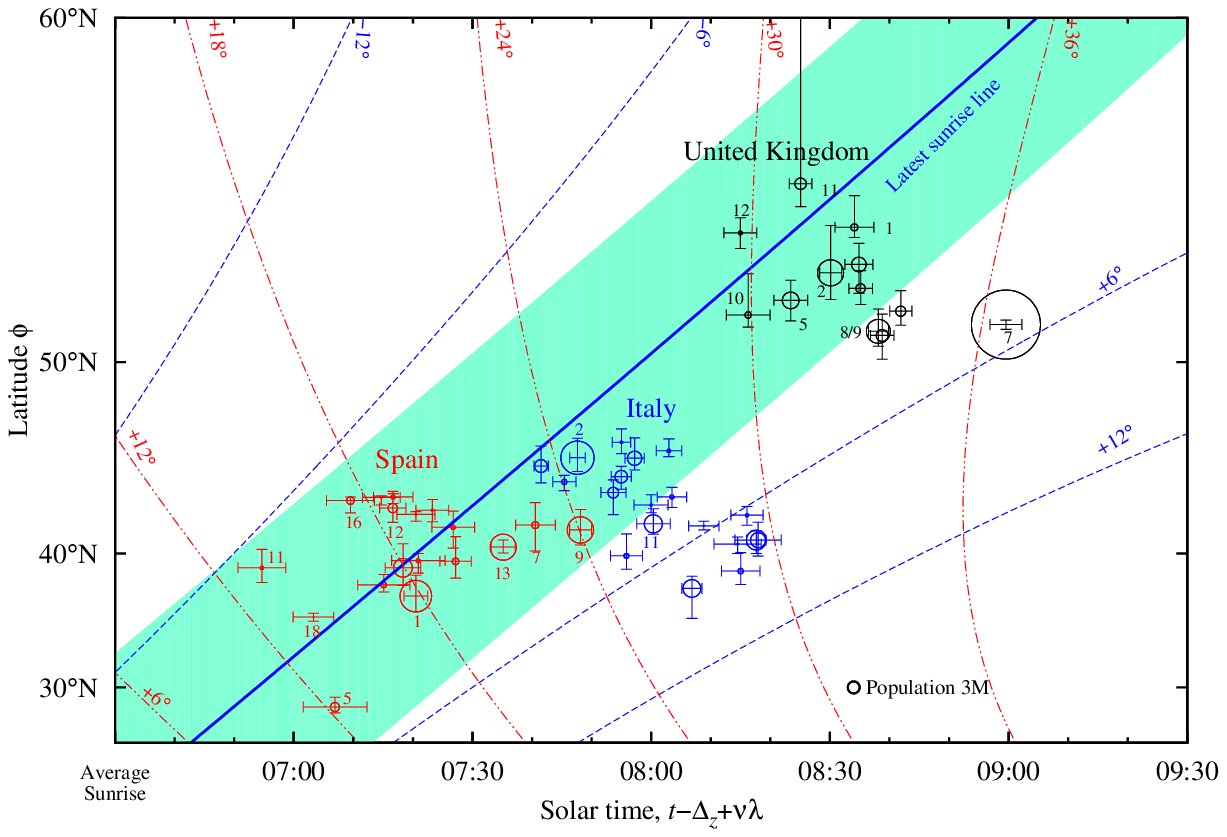}
  \caption{Cuando se considera la ecuación~(\ref{eq:6}) y se transforma $\lambda-\lambda_s$ en hora solar ---véase la ecuación~(\ref{eq:6})--- la familia de rectas de la figura~\ref{fig:extremal} colapsa en una línea (continua, en azul) que representa el amanecer más tardío. La banda de color representa una anchura de media hora antes y después de este amanecer. Los puntos representan el valor mediano de la hora solar de entrada al trabajo de las regiones de España, Italia y Reino Unido. El círculo representa la población de la región. A la derecha de la línea azul es de día en cualquier época del año.  Las líneas discontinuas azules representan diferentes alturas del Sol en el día de solsticio austral. Las líneas discontinuas rojas representan alturas del Sol para el día del solsticio boreal, que representa el extremo opuesto. En cualquier punto el cruce entre una línea roja y una línea azul marca los valores extremos de la posición del Sol a esa hora solar y esa latitud.}
  \label{fig:latitud}
\end{figure*}

La familia de rectas que aparece en la figura~\ref{fig:extremal}  colapsa si en vez de representar en el eje $X$ la longitud geográfica del observador se representa $\lambda-\lambda_s$. Esta magnitud, según la ecuación~(\ref{eq:8}) es una única línea recta cuando se representa frente a la función de la latitud que aparecen en la ecuación.

De la misma forma $\nu(\lambda-\lambda_s)$ determina la distancia temporal al mediodía del amanecer más tardío del año cuando se evalúa la ecuación~(\ref{eq:8}). O, usando la ecuación~(\ref{eq:6}) para $\delta_s=\epsilon$ y $z=0$ obtenemos la hora del orto solar. 

En la figura~\ref{fig:latitud} se representa esta función (línea continua en azul) para la condición más exacta $z=z_0=\unit{-0.83}{\degree}$ que representa el despuntar del limbo superior del Sol. Ahora la función representada no es exactamente una recta pero está muy cercano a serlo, el ojo humano no lo diferencia fácilmente.  La banda de color resalta una distancia temporal de más o menos media hora respecto de este amanecer más tardío.

La figura también muestra diferentes valores de la altura del Sol bajo o sobre el horizonte $z=\left\{\unit{-12}{\degree},\unit{-6}{\degree},\unit{+6}{\degree},\unit{+12}{\degree},\unit{+18}{\degree},\unit{+24}{\degree},\unit{+30}{\degree},\unit{+36}{\degree}\right\}$ tanto para el día del solsticio austral (líneas discontinuas trazadas en azul) como para el día del solsticio boreal (líneas discontinuas con puntos y rayas dibujadas en rojo). Estas líneas no son rectas en la gráfica sino curvas. La curvatura de esas líneas ---debido al valor no nulo de $z$--- muestra indirectamente el ángulo paraláctico: el Sol tarda más tiempo en alcanzar una altura de $\unit{6}{\degree}$ sobre el horizonte según crece la latitud del observador.

Como estás líneas están calculadas para $\delta_s=\pm\epsilon$ son extremales. Si consideramos una línea discontinua azul de altura sobre el horizonte $z_1$, a la derecha de ella el Sol está más alto en cualquier día del año. Si consideramos una línea de puntos y rayas roja de altura sobre el horizonte $z_2$, el Sol está, cualquier día del año, más bajo a su derecha. Cuando las dos líneas se cortan se lee la altura máxima y mínima del Sol en esa latitud y para esa hora solar. Particularmente a la derecha de la línea azul es de día en cualquier época del año. 

Las líneas rojas se han desplazado una hora hacia la derecha para dar cuenta del cambio de hora de marzo. Es decir, el eje $X$ no representa exactamente la hora solar sino la hora solar modificado por este cambio. El cómputo de la condición extrema se ha mantenido en $\delta_s=-\epsilon$ a pesar de que en latitudes por debajo de $\unit{40}{\degree N}$ amanece "<más tarde"> a finales de octubre, justo antes de retrotraer el cambio de hora de marzo.

Obsérvese entonces que la figura~\ref{fig:latitud} muestra en qué medida la hora solar, por si sola, no determina todas la características del Sol en un instante dado.  Una misma hora solar de $\uhora{08:00}$ ---ocho horas después de la medianoche local, cuatro antes del mediodía local--- significa en España una situación en la que el Sol está siempre por encima del horizonte entre $\unit{+6}{\degree}$ (invierno) y $\unit{+24}{\degree}$ (verano). En el Reino Unido, la situación del Sol a esa hora solar es en verano muy similar a la que hay en España, pero en invierno se observa un Sol por debajo del horizonte. Por tanto gráficas y cuadros de valores en las que solo se muestren los valores medianos de los horarios en horas solares, como por ejemplo la figura~\ref{fig:LondonTrabaja}, no contienen toda la información del problema.

Los puntos de la figura~\ref{fig:latitud} representan los valores medianos de la hora solar de entrada al trabajo ---es decir, los valores medianos de la hora civil transformados por la ecuación~(\ref{eq:1}).--- Se observa claramente que estos valores medianos se correlacionan con la hora de amanecer más tardío. La gráfica sugiere que, estadísticamente, los horarios se disponen de forma que las personas medianas inician su actividad próximas a esta línea. En general para la mayoría de las regiones la hora mediana de entrada al trabajo es tal que ocurre siempre después del amanecer sea cual sea la época del año.  Esto es una manifestación directa de cómo el ciclo anual de luz diurna influye estadísticamente en la formación de los horarios.

Al acercarse a esta línea las sociedades modernas, con horarios determinados por el reloj, se acoplan de forma eficiente al ciclo anual de luz. Obviamente en épocas pasadas, antes del uso de luz artificial y del uso del reloj, el inicio de la actividad debía acoplarse directamente con la línea de amanecer, día a día. Ahora es suficiente con mantener el acoplo para la condición extremal invernal, aunque ello suponga desaprovechar parte de la luz solar matutina en los meses de veranos ---el cambio de hora de verano intenta, precisamente, de paliar este efecto.

De acuerdo con esta hipótesis, los horarios más tempranos de inicio de la jornada laboral en España no se deberían al huso "<incorrecto"> sino a la latitud del país, y se producirían para aprovechar al máximo luz solar. Comparado con el Reino Unido, en España amanece una hora antes en el invierno. Comparado con el Reino Unido, en Italia amanece media hora antes en el invierno. Estas diferencias parecen tener influencia en la formación de los horarios con independencia de que los países estén en el huso "<correcto"> (caso de Italia) o en un huso "<incorrecto"> (caso de España). 

Las regiones que se sitúan a la izquierda de la línea de amanecer más tardío quedan suficientemente próximas a ésta; siempre a una distancia angular menor que seis grados por debajo del horizonte. Esta altura solar determina el llamado \emph{crepúsculo civil} que se caracteriza porque el horizonte y los objetos son visibles a simple vista cuando el día es claro. Ninguna región analizada sobrepasa el valor extremo de este crepúsculo civil. En tiempo, el valor mediano nunca va más de veinte minutos antes del amanecer. Esto es muy significativo ya que es un valor de menos de la mitad del incremento que induciría un cambio de huso.

También es importante observar la situación extrema del solsticio boreal determinada por las líneas a rayas y puntos en rojo, donde también se producen diferencias significativas. Las regiones españolas oscilan entre $\unit{+18}{\degree}$ y $\unit{+24}{\degree}$ por encima del horizonte,\footnote{Aquí no se considera el caso de Canarias ya que su latitud casi tropical y el cambio de hora amortiguan enormemente las diferencias entre el verano y el invierno. De hecho la altura del Sol para el valor mediano canario varía entre $\unit{+12}{\degree}$ en verano y $\sim\unit{+3}{\degree}$ en invierno. También dejamos a un lado las regiones de Extremadura y de Ceuta y Melilla por su baja población.} mientras que las italianas lo hacen entre $\unit{24}{\degree}$ y $\unit{30}{\degree}$ y las británicas entre $\unit{30}{\degree}$ y $\unit{36}{\degree}$. Probablemente esto muestre también cómo gestionan las regiones analizadas el problema de la insolación estival y la fría mañana invernal. En unos casos se margina lo primero, en otros casos lo segundo.

En Italia es significativo que las regiones más meridionales entren a trabajar a una hora tan tardía desde el punto de vista solar. Esto se debe a la peculiar estructura horaria de Italia con un valor mediano en hora civil casi constante. Entonces, la hora solar da cuenta de la longitud de la región y en la figura~\ref{fig:latitud} aparece una figura que recuerda a la bota italiana. La figura muestra que este valor constante de hora civil es el menor valor que hace que la entrada al trabajo mediana ocurra después del amanecer, en cualquier época del año y con independencia de la latitud de la región.

. 

La idea poderosa que sugiere la figura~\ref{fig:latitud} es que los horarios de entrada al trabajo se eligen de forma que, estadísticamente, se entre a trabajar después de la salida del Sol con independencia de la época del año. Este es probablemente el condicionador más determinante para el resto de los horarios de inicio de jornada. Sería necesario analizar más países europeos para determinar si esta tendencia es general o no. De momento, en la gráfica analizada aparecen bien ajustados a esta idea tanto la región de Canarias ---en una zona prácticamente subtropical, solo $\unit{5}{\degree}$ al norte del trópico de cáncer--- como la región de Escocia ---$\unit{10}{\degree}$ al sur del círculo polar ártico.--- Razonablemente cabe esperar que este acoplamiento entre el valor mediano de la hora de inicio del trabajo y la línea de amanecer más tardío se quiebre cuando se alcancen latitudes polares, debido a alargamiento de la noche invernal. 

\begin{figure*}[t]
  \centering
  \includegraphics[scale=0.9]{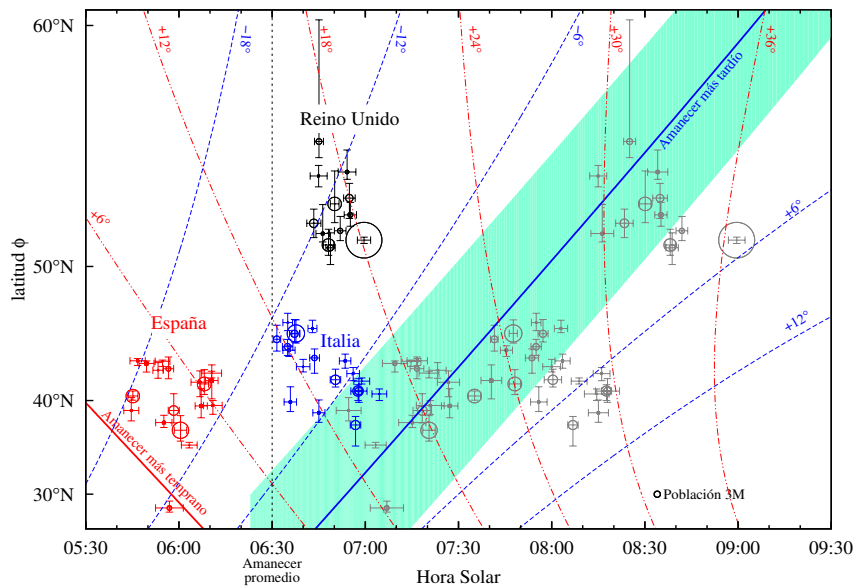}
  \caption{Igual que la figura~\ref{fig:latitud} pero mostrando los valores medianos de entrada al trabajo en gris y los valores medianos de la hora de despertar en rojo (España), azul (Italia) y negro (Reino Unido).}
  \label{fig:latitud-sumer}
\end{figure*}

En la figura~\ref{fig:latitud} también puede estar implícito una competencia entre la situación invernal y la estival. En la figura~\ref{fig:latitud-sumer} se vuelve a mostrar los valores medianos de entrada al trabajo en gris. Además se incluyen los valores medianos de la hora de despertar de las personas que ejercen actividad laboral y se amplía el rango horario para que entre el amanecer estival más temprano.

El desplazamiento hacia valores más tempranos de la hora de inicio del trabajo provoca que la hora de despertar en las regiones españolas sea también más temprano; incluso anteriores a la hora del amanecer promedio. Este desplazamiento acerca la hora de despertar al amanecer del solsticio boreal pero no lo sobrepasa. En resumen podría pensarse que las regiones españolas están cómodas con un horario que les permite entrar a trabajar siempre después del amanecer y que les permite despertarse en verano poco después de la salida del Sol. Es decir, en valor mediano las regiones españolas no madrugan los suficiente como para que siempre sea de noche con independencia de la época del año.

Obsérvese que este esquema es posible porque la diferencia entre el amanecer más temprano y el más tardío en las latitudes de $\unit{40}{\degree N}$ es de aproximadamente dos horas; gracias, también, al cambio de hora que se produce en marzo. Estas dos horas de diferencia coinciden, aproximadamente, con las horas de necesarias para iniciar la actividad laboral: despertarse, cuidados personales, desayunarse, trasladarse al trabajo. Regiones más septentrionales no pueden guiarse por un esquema de este tipo ya que la diferencia entre el amanecer más tardío y el más temprano crece enormemente.

Los datos de fin de jornada de la figura~\ref{fig:LondonTrabaja} no muestran una dependencia con la latitud similar a la que se observa en el inicio de jornada. Esto es una indicación de que el inicio de jornada no está gobernado por las mismas causas que el fin de jornada. 

\begin{figure}
  \centering
\includegraphics[bb=73 82 248 312]{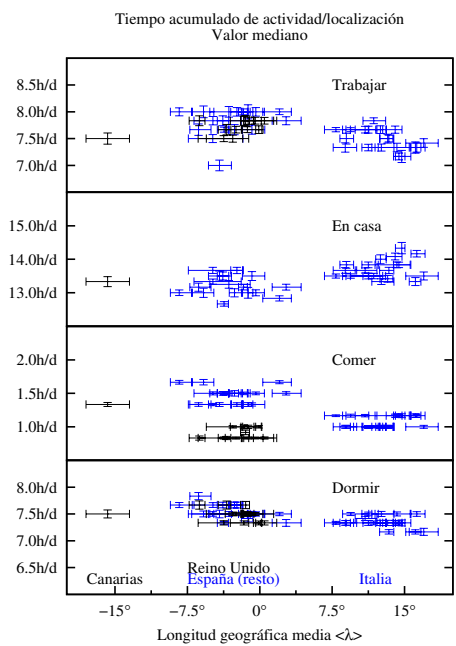}  
  \caption{Valores acumulados de las actividades y localizaciones analizadas para el caso de un día laborable y sobre el campo de personas que trabajan. Obsérvese que el valor mediano del número acumulado de horas que las personas pasan durmiendo o trabajando es muy parecido en las tres encuestas analizadas.}
  \label{fig:acumulatrabaja}
\end{figure}

De la discusión anterior parece deducirse que los trabajadores de las regiones españolas duermen menos y trabajan más que los trabajadores de las regiones italianas y británicas. Sin embargo esto no es así. Los datos acumulados de actividad de los tres países son coincidentes como muestra la figura~\ref{fig:acumulatrabaja}. Esto señalaría que los trabajadores de las regiones españolas típicamente duermen más de una vez, sumando en total una cantidad equivalente a británicos o italianos.

En la sección~\ref{sec:vari-regi-interna} se comentaba que los datos italianos de hora de la cena y de regreso al hogar mostraban un comportamiento anómalo siendo más tardíos cuanto más oriental es la región (véase las figuras~\ref{fig:London} y~\ref{fig:solar}). Puede que este fenómeno esté también influenciado por la latitud. En Italia las regiones más orientales son a la vez las más meridionales, y las regiones más occidentales son más septentrionales: el eje de la bota italiana no se dirige estrictamente al norte sino al noroeste como se observa en las figura~\ref{fig:mapa}. Esto influye necesariamente en el clima y es razonable pensar que esta variabilidad climática induzca la variabilidad del comportamiento humano.

\subsection{{¿}Existe el horario europeo?}
\label{sec:vari-regi-interna}

Podemos imaginar dos modelos extremos de organización horaria. Estos modelos serían los siguientes:

\begin{description}
\item[Modelo 1] Organización uniforme de los horarios a gran escala. Por ejemplo que todas las personas de una determinada zona geográfica relativamente extensa se levantaran a la misma hora porque tienen que entrar a trabajar a la misma hora. En este caso las horas civiles de las actividades serían constantes. Las horas solares tendrían una deriva de pendiente $\nu$ siendo más tempranas en las regiones más occidentales y más tardíos en las regiones más orientales. Las líneas de amanecer más tardío de la figura~\ref{fig:latitud} representan un caso de este tipo: amanece a la misma hora UTC a lo largo de la línea y, consecuentemente, a distinta hora solar.
\item[Modelo 2] Organización de los horarios a nivel local. Las personas se levantan y realizan sus actividades de acuerdo con el Sol. Produciría una situación en la que los horarios solares de las actividades serían constantes y, en contraste, los horarios civiles de dichas actividades tendrían pendiente $-\nu$ siendo más tempranos para las regiones más orientales y más tardíos para las regiones más occidentales. 
\end{description}

Los datos de España, Italia y Reino Unido que aparecen en las figuras~\ref{fig:London} y~\ref{fig:solar} se parecen en conjunto al modelo 2. Esto significa que los horarios de estos países no se fijan uniformemente sino independientemente unos de otros. Esto explica también por qué los horarios españoles ya están adaptados a su meridiano, al contrario de lo que tiende a argumentarse. Que España e Italia compartan el mismo huso horario no implica realmente que en España y en Italia se hagan las mismas cosas a la vez. No hay una autoridad u organización que fije unos horarios comunes en ambos países. O dicho de otra forma, \emph{existe el huso horario central europeo pero no existe "<un horario central europeo">.} 

Este es el fallo esencial de aquellos que asocian el desfase del huso con un presunto \emph{jet lag} permanente.\cite{vidales-jetlag} En esta referencia aparece una fotografía que hizo fortuna en las redes sociales: un mapa de Europa mostrando la línea de la noche ---parte de noche, parte de día---. La línea pasa aproximadamente por Francia y el mar que media entre las islas Baleares y Cerdeña. Se marca la hora civil de Nápoles (donde es de noche y el reloj marca las \uhora{19:00}) y la hora civil de Castellón (donde aún hay luz solar y el reloj  \uhora{19:00}).

No hay ningún problema con esa figura porque \emph{los ciudadanos de Castellón no están realizando las mismas actividades que los ciudadanos de Nápoles}. Por eso el reloj puede marcar lo mismo sin que ninguno de ellos tenga \emph{jet lag}.  Obviamente los napolitanos harán las cosas antes ---cronológicamente--- que los castellonenses, como muestra la figura~\ref{fig:London}, al estar más hacia oriente. Pero, como muestra también la figura~\ref{fig:London}, el desfase ---el \emph{lag}--- coincide justamente con el desfase solar debido a la rotación de la Tierra. Es decir, no hay ninguna alteración significativa.

A nivel interno cada país muestra estructuras horarias diferentes. Así España y Reino Unido muestran internamente comportamientos cercanos al modelo 2. Esto quiere decir que en buena medida sus horarios están fijados a nivel intrarregional más que a nivel nacional. 

El caso italiano es más disperso en este sentido. Los datos de inicio de jornada se parecen a los del modelo 1. Los datos de fin de jornada, también; con excepción de hora de cenar y la hora de regreso al hogar que muestran una tendencia diferente a estos modelos. Aquí, los datos italianos muestran una pendiente contraria a la evolución solar siendo más tardíos cuanto más oriental es la región. Pero por la estructura geográfica italiana las regiones más orientales son, a la vez, las más meridionales y las regiones más occidentales son las más septentrionales así que este fenómeno puede ser debido a la latitud.

Otra posible explicación de esta diferencia de comportamiento entre los datos italianos y británicos o españoles puede ser la descentralización de los países. Es razonable que un país fuertemente centralizado y relativamente poco amplio geográficamente pueda producir una estructura horaria similar al modelo 1. Mientras que un país descentralizado o suficientemente amplio geográficamente tienda más al modelo 2.

\subsection{Los tard{\'\i}os horarios de la televisi{\'o}n espa{\~n}ola y el f{\'u}tbol espa{\~n}ol}
\label{sec:los-tardios-horarios}

Un tema recurrente en la discusión actual es que los horarios de la televisión española ---el conocido ahora como \emph{prime time}--- son muy tardíos lo que ocasiona que los españoles se acuesten muy tarde.\cite{aavv-primetime} Las televisiones españolas argumentan que el horario de \emph{prime time} está relacionado con la presencia de ciudadanos en los hogares y que esta presencia es más tardía por lo que los horarios de \emph{prime time} son más tardíos.

Los datos presentados en la Sección~\ref{sec:resultados}, particularmente la figura~\ref{fig:Civil} concuerdan con el argumento de las televisiones. Además hay que matizar que los resultados muestran también que españoles no regresan a casa más tarde que los italianos por culpa del huso vigente sino, exclusivamente, porque están están situados más hacia occidente. En horas solares los españoles regresan incluso un poco antes que los italianos. Los datos no avalan afirmaciones como esta:
\begin{quote}
  "<Los datos son espectaculares: en España, a las ocho de la tarde, solo el 50\% [el valor mediano] de la gente está ya en casa y hasta las diez no están el 80\%. Las consecuencias son desastrosas.">\cite{gabriela-comer}
\end{quote}
salvo que también sean espectaculares y desastrosos en Italia y Reino Unido.

Como no hay nada anómalo en el horario de regreso al hogar de los españoles no hay nada anómalo en el horario televisivo español. Este, también, se ha adaptado al huso vigente y a la actividad solar local. Por eso afirmaciones como\cite{buqueras-tv}
\begin{quote}
  Estos horarios [los del \emph{prime time} televisivo español] tan anómalos crean problemas a muchas personas, ya que, al reducir las horas de sueño, aumentan el cansancio y el estrés, propician una mayor siniestralidad laboral y de tráfico, así como el absentismo, y afectan a la productividad y al rendimiento escolar. 
\end{quote}
no encuentran justificación en los datos de las encuestas analizadas.

O, como se ha señalado ya anteriormente, que España ---excepto Canarias--- tenga el mismo huso horario que Italia no implica que los ciudadanos españoles hagan las cosas a la vez que Italia. Así, si un noticiario empieza en Italia a las \uhora{20:00}, es lógico que en España empiece a las \uhora{21:00}, precisamente porque ambos países tienen el mismo huso horario. Esa hora de diferencia da cuenta de la diferencia de longitud geográfica entre ambos países.\footnote{Si uno se toma la molestia de corregir por longitud los valores del \emph{prime time} que aparecen en la referencia~\onlinecite{aavv-primetime} observará que los horarios italianos y españoles son similares.}

De la misma forma cabe analizar los horarios de fútbol. La hora de inicio de los encuentros de fútbol de la  \textsc{UEFA}\footnote{Acrónimo de "<Union des associations européennes de football">, la entidad organizativa de competiciones de fútbol asociación a nivel europeo.} \emph{Champions League} es las \uhora{20:45CET}, con independencia del lugar en el que se celebre el encuentro.\footnote{Salvo que se dispute en Rusia} Esto es un ejemplo de organización horaria del modelo 1 señalado en la sección~\ref{sec:vari-regi-interna} con una organización fijando un horario en una escala geográfica amplia. 

Esa hora de inicio es, aproximadamente, una hora más tarde que la hora mediana de regreso al hogar en Italia y coincide con el \emph{prime time} italiano. Por contra coincide con el valor mediano de la hora de regreso al hogar en España y se adelanta al \emph{prime time} español. 

Por eso, cuando cada uno de esos países organiza competiciones propias ocurre que la hora de inicio de los encuentros italianos nocturnos coincide las \uhora{20:45CET} mientras que en España se retrasan una hora. Esta hora vuelve a ser la hora de diferencia solar que media entre un país y el otro.

Dicho de otra forma los horarios solares de inicio de los partidos nocturnos italianos cuyo horario se fija localmente ---típicamente \uhora{21:00CET} en Italia y \uhora{22:00CET} en España--- coinciden y la diferencia en horas civiles responde a la diferencia de longitud.

\subsection{La amplitud geogr{\'a}fica: el caso de Galicia}
\label{sec:la-ampl-geogr}

El caso de Galicia, la región más occidental de la España peninsular, es también objeto de debate en la discusión de los horarios españoles. Incluso  algunos piden un huso horario diferente al del resto de la España peninsular.\cite{horariogallego-lavoz}

Este problema ha de relacionarse con la amplitud geográfica de la España peninsular que, incluyendo a las Islas Baleares alcanza los $\unit{12}{\degree}$ ---casi una hora solar de diferencia---. Los resultados de este informe no muestran que Galicia esté especialmente perjudicada por esta cuestión. Tanto la figura~\ref{fig:London} como las figura~\ref{fig:solar} muestran que Galicia tiene unos valores de las actividades que si bien son tempranos no cabe señalarlos como anómalamente tempranos. La figura~\ref{fig:latitud} a la izquierda muestra a Galicia (12) dentro de la banda anormalmente temprana que se adentra en la región de trabajos antes del amanecer. Sin embargo no es la región que peor se comporta en este aspecto a pesar de ser la más occidental. Significativamente País Vasco (16) obtiene un resultado aún más desplazado.

A este respecto hay que considerar lo que ya se ha discutido en la sección~\ref{sec:vari-regi-interna}. En la medida en que existen horarios civiles fijados a nivel suprarregional es lógico que las regiones más occidentales tiendan a valores solares más tempranos. Que los horarios solares gallegos estén en unos valores tempranos, pero no anómalos, es por tanto razonable y apunta a dos factores: primero que la diferencia de longitud no es demasiado grande, de forma que horarios fijados a nivel suprarregional no producen en Galicia horas solares demasiado tempranas.\footnote{En el ejemplo ya señalado del fútbol europeo hablábamos de una hora, las \uhora{20:45CET}, fijada a escala europea que, expresada en hora solar, varía hasta dos horas entre Polonia y Galicia, por ejemplo. Los horarios fijados a escala española solo producen variaciones de una hora solar dentro de la península ibérica.} Segundo que los horarios gallegos están, también en buena parte, fijados en base a decisiones regionales internas y adecuados a su posición geográfica. Cabe pensar que este proceso de adecuación se haya ido llevando a cabo desde el cambio horario del año  1940 y, también, desde la descentralización del poder acontecida en el año 1978.

\subsection{El verdadero problema del huso horario espa{\~n}ol}
\label{sec:el-problema-del}

Dando por sentado que las horas solares de actividad española implican que sus horarios civiles están acompasados correctamente con su meridiano voy a abordar dónde reside realmente el problema de horario español. Este problema es realmente pequeño, más bien anecdótico.

El problema está ---sigue estando---  cuando comparamos ---y seguimos comparando--- horas civiles, que es la magnitud que normalmente se compara cuando analizamos datos de España y del resto del mundo. Véase por ejemplo las referencias Ref.~[\onlinecite{vidales-jetlag}] o~[\onlinecite{yardley-dinners}]. 

El ejemplo más común de este tipo de comparaciones/equivocaciones ocurre cuando viajamos. Bien cuando viajan españoles al resto del mundo, bien cuando extranjeros viajan a España. Tomemos el caso, por ejemplo, de un ciudadano que viaja de Londres a Castellón y, religiosamente, adelanta su reloj de WET (UTC+00) a CET (UTC+01) \emph{a pesar de que no ha cambiado su meridiano}. Cuando su reloj alcanza la hora mediana de comer toma la decisión de comer. El problema es que ahora lo está haciendo una hora solar completa más tempranamente que cuando comía en el Reino Unido puesto que adelantó el reloj pero no se movió de meridiano. Lo hará, también, más tempranamente que los lugareños. Si este ciudadano hubiera sido Phileas Fogg en 1873 no habría cambiado su reloj y, sin duda, habría comido a una hora "<normal"> para la ciudad en la que se encuentra.

Igualmente un ciudadano italiano viaja de Nápoles a Castellón y que, en cambio, no modifica la hora su reloj. Cuando mira su reloj y ve que ha alcanzado la hora mediana de almorzar toma la decisión de comer. También lo hará una hora solar completa antes de lo que lo hacía en la ciudad de origen. Si ese ciudadano se hubiera comportado como Phileas Fogg habría retrasado su reloj una hora y habría comido en paz con los lugareños. 

No debemos desdeñar los efectos positivos de esta peculiaridad: es esto lo que hace que los restaurantes de zonas turísticas tengan una clientela muy bien distribuida durante muchas horas. Es un coste de oportunidad y una ventaja económica que primero coman turistas extranjeros y después coman indígenas españoles escalonadamente en vez de que se agolpen todos en un lapso de tiempo pequeño. 

El mismo efecto ocurre en sentido contrario si un ciudadano español viaja al extranjero. Recientemente un periodista español destacado en Varsovia (Polonia) comentaba los horarios locales\cite{suanzes-tuiter-20140528} y los listaba de la forma que aparece en la segunda columna del cuadro~\ref{tab:polonia}. Esos valores, expresados en horas civiles, se destacan como tempranos. El horario civil es el mismo en Varsovia y en Madrid pero el periodista no cuenta con el "<jetlag"> que sí ha experimentado. Al trasladarse de Madrid a Varsovia se desplaza $\unit{24.7}{\degree}$ hacia el este. Siguiendo la costumbre de Phileas Fogg debió adelantar su reloj en una hora y treinta y nueve minutos. Si lo hubiera hecho su reloj marcaría las horas de la tercera columna y no consideraría esos horarios como adelantados salvo el de la cena cuya variación puede estar inducida por el cambio de latitud.

\begin{table}
  \centering
  \begin{tabular}{lcc}
    \toprule
    \textbf{Nombre}&\textbf{Hora civil}&\textbf{Hora equivalente}\\
&&\textbf{a Madrid}\\
\colrule
Primer desayuno&$\uhora{07:45}$&$\uhora{09:24}$\\
Segundo desayuno&$\uhora{09:00}$&$\uhora{10:39}$\\
Tercer desayuno&$\uhora{11:00}$&$\uhora{12:39}$\\
Comida&$\uhora{13:30}$&$\uhora{15:09}$\\
Cena&$\uhora{18:15}$&$\uhora{19:54}$\\
\botrule
  \end{tabular}
  \caption{Ejemplo de usos horarios en Polonia.\cite{suanzes-tuiter-20140528} En la segunda columna los valores en hora civil. En la tercera columna los mismos valores con el añadido del "<jetlag"> correspondiente a trasladarse desde el meridiano de Madrid ($\unit{3.72}{\degree W}$) al de Varsovia ($\unit{21.02}{\degree E}$) y que equivale a un adelanto de \uhora{01:39} horas según la razón $\nu$. }
  \label{tab:polonia}
\end{table}

Indirectamente el cuadro~\ref{tab:polonia} muestra, a modo de ejemplo sin relevancia estadística alguna, que los horarios españoles no son más tardíos que los polacos. Al contrario, parecen más tempranos.

A veces esta peculiaridad de los horarios españoles provoca decisiones completamente erróneas.\cite{arribas-osaka-maraton} En el año 2007 los campeonatos mundiales de atletismo se disputaron en Osaka (Japón). El inicio de la prueba de la maratón fue fijado a las hora local \uhora{07:00UTC+09}. Un atleta español, José Ríos, adecuó su entrenamiento a las condiciones de la prueba y entrenó diariamente a las \uhora{07:00} hora civil local en Santander y Barcelona. José Ríos sufría como pocos los efectos de la humedad ambiental. Había estado en Japón e incluso había ganados dos maratones allí. Nada de ello le sirvió para darse cuenta de que a las \uhora{07:00} hora local de Osaka el Sol ya brilla en el cielo mientras que el empezaba a entrenar incluso antes de que saliera el Sol.\footnote{El Sol salió a las \uhora{05:24UTC+09} en Osaka el día de la prueba. Una hora y treinta y seis minutos antes del inicio de la prueba. A la hora de este inicio tenía ya una altura considerable. Ese mismo día amaneció a las \uhora{07:30} hora local en Santander ---media hora después de la hora a la que entrenaba el atleta---  y a las \uhora{07:09} en Barcelona. Una hora de estas dos horas de diferencia se debe a que España observa horario de verano mientras que Japón no. La otra es debida a la peculariedad del huso español.} Siendo la humedad un factor preocupante para el atleta y siendo que esta propiedad varía especialmente tras el amanecer, la preparación fue poco afortunada.

En sentido opuesto cabe valorar la decisión del Congreso de los Diputados de iniciar ciertas sesiones a las ocho de la mañana ---en vez de la tradicional hora de las nueve de la mañana--- para conciliar la vida laboral y familiar.\cite{marin-congreso-ocho} Teniendo en cuenta que, en general, ya realizamos las actividades de inicio de jornada en el rango de lo temprano, la decisión suponía desplazarse más en el sentido de la (pequeña) anomalía. Estas sesiones se iniciaron en la primavera del 2006; en otoño e invierno tal horario habría hecho comenzar las sesiones antes de que saliera el Sol a una hora mucho más temprana de la que habitualmente se observa en otros parlamentos.

\subsection{{¿}Por qu{\'e} los horarios espa{\~n}oles ya est{\'a}n acoplados con su meridiano?}
\label{sec:el-trasf-hist}

Las autoridades tienen dos formas de actuar sobre los horarios. La primera es modificándolos directamente fijando las horas de inicio de las actividades que sean de su competencia. La segunda es modificando el huso horario.

Puede trazarse un símil económico: modificar el huso horario es algo parecido a "<devaluar ---o revalorizar--- una moneda">. Modificar los horarios vigentes es decretar una "<subida o bajada de los sueldos">, también llamado en estos tiempo "<devaluación interna">. 

Igual que en economía las autoridades prefieren actuar sobre los husos horarios desde que en el siglo \textsc{XX} se instituyó el sistema de husos. Es lo que se hace todos los años cuando en marzo se establece el horario de verano. Es mucho más fácil decretar el cambio de huso que decretar un cambio de todos los horarios. Además, el cambio de huso parece aséptico ---"<debe usted adelantar la hora">--- mientras que el cambio de horario no, ya que habría que hacerse mediante alguna fórmula que decretara "<madrugar más"> al día siguiente.

Al adoptar esta estrategia las autoridades dejan amplio margen de actuación a las personas para fijar los horarios dentro del huso vigente. Esto es la causa de que los horarios españoles ya estén acoplados con la actividad solar y es el parámetro que pasa desapercibido al discutir el tema.

Podemos imaginar que a principios del siglo pasado los horarios españoles estaban acoplados con el huso de forma que la figura~\ref{fig:latitud} tenía un aspecto similar. Las personas empezaban a trabajar siempre de día. Necesariamente eso implica que se pierde parte de la mañana en el amanecer de los días de verano especialmente si, como ocurría entonces, no se hacía un cambio de hora en marzo.  En el año 1940 tras la Guerra Civil la situación debía ser similar dentro del contexto caótico que debía vivirse en esos momentos. Entonces se realiza el cambio de huso\footnote{Es conveniente aclarar que el cambio de huso se realizó inicialmente en el año 1938 en la zona controlada por la Segunda República. Con este huso murió en abril del año 1939. Un año después el régimen de Franco readoptó este huso en todo el territorio español.\cite{planesas-ign-2013}} adelantándolo una hora con la idea de hacer madrugar a las personas y "<aprovechar las mañanas">. 

En la figura~\ref{fig:latitud} esto significa que los horarios se desplazan una hora hacia la izquierda. Probablemente los horarios entran demasiado en la zona de sombra en la que es de noche durante una época del año. Cabe pensar razonablemente que se dispara entonces un proceso que lleva los horarios a una situación similar a la anterior. Es imposible dar detalles del proceso ya que no hay registros pero no hay alternativa posible: o lo horarios españoles estaban desfasados en el año 1940 y el cambio de huso los puso en fase mágicamente o, el cambio de huso del año 1940 desfasó los horarios y, con el tiempo, volvieron a quedar acoplados con la actividad solar. Es razonable pensar también que en el contexto socioeconómico de la época, con la actividad económica prácticamente paralizada, este cambio fuera relativamente fácil: la actividad socioeconómica nueva nacía ya adaptada al nuevo huso.

En realidad la mejor muestra del éxito del proceso adaptativo es la propia pervivencia del huso horario.\footnote{Como ocurre también en Francia, Bélgica y en Holanda; en contraste con los fracasos de cambio de huso habidos en Reino Unido y Portugal en el año 1968 y en el 1992 en una situación socioeconómica muy diferente.} O bien los horarios estaban desfasados y el cambio fue a mejor, o los procesos que se dispararon tras el cambio fueron posibles y rápidos. Lo cierto es que aunque el decreto de cambio de hora indicaba que "<oportunamente se señalará la fecha en que haya que restablecerse la hora normal[sic]">, la fecha nunca llegó. La adaptación, sí.

Seguro que pueden ponerse muchos ejemplos de estos cambios pero voy a traer solamente dos. Primero el caso de las corridas de toros, que es una actividad singular porque está íntimamente ligada al Sol ya que los precios de venta de las entradas varían según la localidad asignada reciba insolación o no. Esto significa que los horarios solares de las corridas de toros no pueden variar significativamente. Así hemos pasado de las cinco de la tarde lorquianas, que eran las \uhora{17:00UTC+00}, a las más actuales \uhora{18:30UTC+02} o \uhora{19:30UTC+02}. En realidad, las tres horas son similares ya que las actuales equivalen a las \uhora{16:30UTC+00} y \uhora{17:30UTC+00}, respectivamente. Este cambio tuvo que realizarse inmediatamente al cambio de huso horario dada las características de la actividad.

El otro ejemplo tiene que ver con los horarios de las votaciones. Esta no es una actividad directamente ligada al Sol pero sí indirectamente: a ninguna administración se le ocurre iniciar unas votaciones a unas horas demasiado tempranas como para que no haya amanecido. Recientemente el escritor Xavier Pericay describía\cite{pericay-aleman} las últimas elecciones al parlamento alemán de la República de Weimar celebradas el 5 de marzo del año 1933 con las siguientes palabras:
\begin{quote}
  Aquel día, que también era domingo, los residentes alemanes con derecho a voto fueron convocados en el muelle de San Beltrán del puerto barcelonés, donde está hoy la terminal de cruceros. En dos turnos: unos tenían cita a las siete y cuarto de la mañana; otros, a la una y media de la tarde –un horario muy poco español, por cierto–
\end{quote}
Las \uhora{07:15} del 1933 eran las \uhora{07:15UTC+00}. Hoy serían las \uhora{08:15UTC+01} que no es muy diferente de las horas actuales de inicio de las votaciones.\footnote{La hora actual de inicio de las votaciones es las nueve de la mañana; ligeramente más tarde que la hora mediana de entrada al trabajo de las regiones españolas. Su situación en la figura~\ref{fig:latitud} sería similar a los datos de esta variable.}

En este sentido el efecto que se busca al realizar un cambio horario como el que se hizo en el año 1940 ---hacer madrugar a los habitantes--- se va difuminando a lo largo del tiempo cuando los horarios vuelven a converger con la actividad solar media. 

No es menos importante en esta discusión el hecho de que no exista un horario europeo ---véase la sección~\ref{sec:vari-regi-interna}. Cuando se usa la expresión "<en España rige la hora de Berlín"> no se puede querer decir que los horarios de las actividades de las regiones españoles estén fijados desde "<Berlín">. Y como efectivamente los horarios se fijan a escala nacional ---véase la figura~\ref{fig:London}---, los horarios españoles se han adecuado con su meridiano, tal y como hacen los horarios italianos con el suyo.

\section{Propuestas de actuaci{\'o}n}
\label{sec:concl-prop-de}

La conclusión de este trabajo es que los horarios españoles son comparables a los italianos o los británicos. Al menos en las encuestas de uso de tiempo no hay nada que destaque significativamente en los horarios españoles respecto de los otros dos analizados. Claro que para llegar a esta conclusión uno no puede simplemente analizar los valores medianos en la hora civil del lugar.

La figura~\ref{fig:London} o la figura~\ref{fig:solar} muestra que los horarios de los tres países están correlacionados por el movimiento aparente del Sol de forma natural realizando las actividades de inicio y fin de jornada en horas comparables a los países analizados. Es decir, los españoles no viven en ningún tipo de \emph{jet lag} permanente por usar el huso horario de Europa central. Simplemente es la región más occidental con dicho huso. 

La figura~\ref{fig:latitud} sugiere que los horarios de entrada al trabajo de los españoles son más tempranos que los italianos y británicos por la menor latitud de España. 

 Tampoco parece que el tiempo total que dedican los ciudadanos españoles a dormir o a trabajar difiera mucho del de los italianos o británicos ---véase las figuras~\ref{fig:acumulado} y~\ref{fig:acumulatrabaja},--- en contra de lo que habitualmente se sostiene.\cite{estevil-duermepoco,bolano-diferent,fernandezcrehuet-fedea,buqueras-tv}

Cualquier actuación sobre el huso horario debería ser extremadamente cautelosa para  no romper la tendencia general que muestran estas figuras. La cautela es importante porque no pensamos cambiar el huso horario de forma continua sino a intervalos discretos: o estamos en el huso CVT, o en el WET o en el CET,  nadie imagina una solución a la venezolana o a la iraní trasladando medio huso, o a la nepalí, con un cuarto de huso.

Aunque lo que trasciende de los comentarios que se hacen públicos es que los horarios españoles son tremendamente anómalos por tardíos y que el origen de todos los males están en el huso horario, a la hora de las propuestas las cautelas se hacen presentes. Por ejemplo, el Informe de la Subcomisión creada en el seno de la Comisión de Igualdad para el Estudio de la Racionalización de Horarios, la Conciliación de la Vida Personal, Familiar y Laboral y la Corresponsabilidad decía en su informe final citando a Jos Colin:\footnote{Jos Colin era colaborador externo del Centro Internacional Trabajo y Familia del IESE Business School quien compareció por delegación de Nuria Chinchilla directora del Centro}
\begin{quote}
  "<volver al huso horario europeo occidental vigente antes de la guerra (...). Ello iría unido al adelanto en una hora de todos los horarios de referencia social (programas de televisión, partidos de fútbol, misas) a partir de la mitad de la mañana.">\cite{informe-subcomision} (página~26)
\end{quote}

Esto es una actuación mixta que decreta el cambio de huso ---la "<devaluación">--- y el adelanto de los horarios vespertinos ---la "<devaluación interna">. Este cambio mixto no implica nada para el segmento de tarde donde se producen los horarios de fin de jornada. Efectivamente, el retraso del huso y el adelanto de los horarios se neutralizan: si el noticiario empezaba a las \uhora{21:00CET}, luego empezaría a las \uhora{20:00WET}. Solo que ambos dos instantes de tiempo son exactamente el mismo.

De forma explícita esto es un reconocimiento de que los horarios vespertinos españoles ya funcionan perfectamente y que no es necesario cambiar nada. Una idea que no se transmite a la opinión publica española.

La propuesta, sin embargo, sí actúa sobre los horarios matinales ya que en ese segmento solo se produce un retraso del huso y, por tanto, un retraso de los horarios. Probablemente este cambio está pensado en alterar la tendencia de los horarios de entrada al trabajo, que son más tempranos de lo que se observan en Italia y Reino Unido.

\begin{figure}
  \centering
    \includegraphics[scale=0.6]{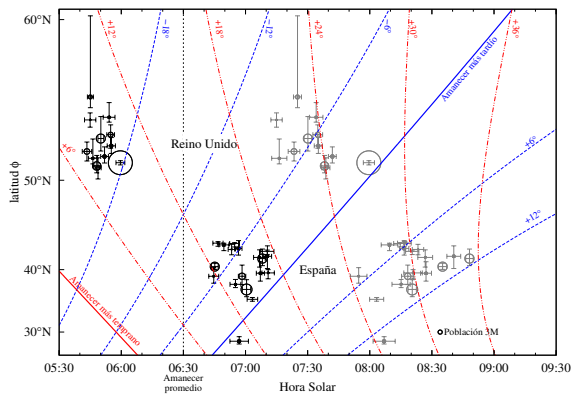}
  \caption{Simulación de lo que ocurriría si España retrasa el huso una hora y, consecuentemente, los horarios de la figura~\ref{fig:latitud-sumer} se desplazan una hora hacia la derecha. Y simulación de lo que ocurriría en Reino Unido si se adelantara el huso una hora, como se hizo en 1968. Los horarios se desplazan una hora hacia la izquierda. En negro los horarios medianos de despertar; en gris los horarios medianos de entrada al trabajo. Las bandas representan incrementos de media hora desde la la línea del amanecer más tardío.}
  \label{fig:latitud-2}
\end{figure}

La figura~\ref{fig:latitud} muestra que esta diferencia puede tener un origen natural. Además, es una excelente plataforma para simular los efectos que tendría dicho cambio. En la figura el retraso del huso sería equivalente a trasladar hacia la derecha los valores medianos; un adelanto del huso sería equivalente a desplazar hacia la izquierda los valores medianos. En la figura~\ref{fig:latitud-2} se muestra una simulación de lo que ocurriría si se retrasase el huso de España y, también, si se adelantase el huso de Reino Unido ---tal y como ocurrió en 1968.  

En el caso español, el retraso del huso genera, inmediatamente, un hueco horario en la que el Sol está plenamente visible pero no se ha alcanzado el valor mediano de la hora de inicio del trabajo. Incluso en el día de más tardío amanecer habría regiones españolas para las cuales este valor excedería en más de una hora la salida del Sol.  

Además, si en la situación actual el Sol tiene una altura sobre el horizonte de  entre $z=\unit{+18}{\degree}$ y $z=\unit{+24}{\degree}$ a la hora mediana de entrada al trabajo; la hora de retraso que induce el cambio de huso le permite al Sol alcanzar una altura de entre $z=\unit{+30}{\degree}$ y $z=\unit{+36}{\degree}$ en verano.\footnote{Esta situación se produciría igualmente si se cancelara el cambio de hora de marzo.} Esto es una gran diferencia para el seno cuadrado del ángulo $z$ ---que es el parámetro que determina los fenómenos asociados a la insolación--- que pasaría del intervalo $(0.0954,0.165)$ al intervalo $(0.250,0.345)$; doblando los valores con el cambio de huso. Esta alteración sería especialmente significativo para las personas que realizan actividad física en el exterior por las implicaciones que tendría para su salud.

La figura~\ref{fig:latitud-2} también muestra una simulación de qué ocurriría si el horario del Reino Unido pasara de WET a CET, que es el movimiento contrario. Este cambio se realizó en 1968 y fracasó en gran medida porque, como se observa en la figura, movió los horarios a una zona anómala ya que, en invierno, las personas debían entrar a trabajar bastante antes de que hubiera amanecido. La hora de despertar invernal se situaría por debajo del crepúsculo astronómico ($z<\unit{-18}{\degree}$), ya en la noche profunda.

En vez de el cambio de huso se podría fomentar que las regiones que muestran una entrada al trabajo más temprana que la hora más temprana de amanecer corrigieran la tendencia. El primer paso para hacer esto es ser consciente de esta pequeña anomalía.

Una propuesta diferente sería un doble cambio como el que se ha citado pero obviando los "<horarios de referencia sociales"> y la distinción de "<a partir de la mitad de la mañana"> ---habría que determinar qué significa esto.--- Es decir, declarar un retardo del huso trasladándolo de CET a WET ---y de WET a CVT en Canarias--- y un adelanto de los horarios declarando que si antes usted entraba a trabajar a las \uhora{08:00} civil, ahora pasará a entrar a trabajar a las \uhora{07:00} civil. El doble cambio ---retraso y adelanto--- implicaría dejar la figura~\ref{fig:London} y la figura~\ref{fig:latitud} inalteradas ya que las antiguas \uhora{08:00CET} serían coincidentes con las nuevas \uhora{07:00WET}. 

Con ese cambio, efectivamente, retornaríamos al huso horario natural y el mediodía solar coincidiría ---en Castellón--- con el mediodía civil ---al menos en invierno.--- Esto, sin embargo, no es ninguna de una ventaja: se trata puramente de una cuestión sentimental sin relevancia alguna.

También así los horarios civiles españoles se podrían comparar con los del resto de Europa de forma más directa. Pero esto no es tampoco una ventaja adicional para el ciudadano común. 

También observaríamos que las horas de las comidas de los turistas coincidirían con la de los españoles. Es decir, los restaurantes de las zonas turísticas recibirían los mismos comensales ---indígenas y forasteros--- en un lapso de tiempo menor.

La consecuencia más importante de este cambio sería, sin embargo, de tipo psicológico. El cambio sería, por estos efectos, muy parecido a un cambio de moneda como el que se efectuó en 2002 con la entrada del euro. Estamos acostumbrados a unos horarios para las las actividades relacionados con el Sol; esta costumbre fija unos niveles de "<temprano"> y "<tarde"> que están basadas en el horario actual. Cambiarlo implica trastocar estas equivalencias: en algún momento habría que explicar al ciudadano que ese cambio fantástico que va a llevar el mediodía a las doce y que va a adelantar el telediario a las \uhora{20:00} como en Europa implica, también, levantarse a las $\uhora{06:00}$. Dejaría a la gente ---que está acostumbrada psicológicamente a manejarse en un horario civil CET viendo al oeste del meridiano principal--- viviendo en otro horario.

\section{Conclusi{ó}n}
\label{sec:conclusion}

Si usted es un ciudadano mediano español ---ese que no existe--- se habrá visto reconocido en el estudio: usted se levanta buena parte del año a oscuras pero en otra época del año verá salir el Sol. Usted entrará a trabajar de día sea cual sea la época del año. Lo que este estudio muestra es que usted no es un ser extraño que viva en un \emph{jetlag} permanente. Al menos los italianos y los británicos medianos se comportan como usted: salen del trabajo, regresan al hogar, cenan y se acuestan a horas parecidas. La pregunta que debe hacerse es ¿qué pretende mejorarse con un cambio de huso?

No es solo el reloj quien rige los horarios porque estos no se eligen aleatoriamente de entre todas las horas disponibles. Tampoco se elige aleatoriamente una hora de inicio de la actividad de forma que el resto de actividades quede ordenada cronológicamente\footnote{Como, analógicamente, puede sortearse una letra o una fecha y, a partir de ella, iniciar una secuencia alfabética o cronológica de llamamientos.}. En este problema inicial el Sol sigue teniendo una influencia directa y determina gravemente la respuesta estadística del conjunto analizado.  

En este sentido las autoridades pueden cambiar el huso las veces que quieran: o bien el cambio será un fracaso ---por impopular--- o bien los horarios se adaptarán lentamente al nuevo huso. 

\def\acknowledgmentsname{Agradecimientos}
\acknowledgments

Tengo un mol de agradecimientos por repartir. Quisiera expresar mi agradecimiento a las instituciones que han realizado las encuestas de uso de tiempo que se han analizado en este estudio. No solo por su realización sino por ponerlas a disposición de los investigadores. Y porque, en general, tienen una instrucciones suficientemente claras como para que un neófito en la materia pueda sacar abundante informaicón.

Todo el informe se ha realizado con herramientas informáticas de libre distribución. Bueno, todo no... \texttt{google} ha sido de mucha ayuda. 

Los datos de las encuestas se han tratado con un programa realizado en lenguaje \textsc{C} y compilado con \texttt{gcc}. Los datos se han volcado posteriormente en \texttt{octave} para realizar estadísticas y cálculos. Las gráficas se han realizado con \texttt{gnuplot}\cite{gnuplot} y el texto se ha compuesto con REV\TeX\ sobre \texttt{GNUemacs} y \texttt{AUC}\TeX.

Ningún animal ha resultado herido o ha sufrido maltrato en la realización de este estudio; autor incluido.

Los datos geográficos se han sacado de \texttt{geonames}\cite{geonames}. Los datos de los mapas poligonales se han obtenido de la página de Hagen Wierstorf\cite{wierstorf-gnuplotting} quien, a su vez, las extrajo de \texttt{Naturalearth}\cite{naturalearth}.

\end{document}


\author{José María Martín Olalla}
\email{olalla@us.es}
\homepage{@MartinOlalla\_JM}
\affiliation{Departamento de F{\'\i}sica de la Materia Condensada. Universidad de Sevilla. PO Box 1065, ES41080. Sevilla. Spain}
\title{Is Spanish time schedule really weird?}
\keywords{time zone; labor time; solar time; time use survey;italy;united kingdom}
\published{June 19th, 2014}
\begin{abstract}
Time use surveys in Spain, Italy and United Kingdom are analysed to provide median times for start of day and end of day activities. These median times are shown to follow the regular pattern provided by the solar apparent motion in the sky. Furthermore, labour start time are revealed to increase with latitude and it is shown that this tendency seems coupled to the latest sunrise time, which is a function of latitude.

Spanish timetables are then shown to be consistent with British and Italian timetables within the accuracy of half a time zone if solar motion is taken into account despite the fact that Spanish timetable ---when expressed in local time--- looks weird.
\end{abstract}
\maketitle

In the past few months Spanish social media have developed lot of pieces on the Spanish time schedule, which is characterised by late times.\cite{vidales-outoftime} The issue has caught attention of the Legislative\cite{informe-subcomision} and international media as well.\cite{bbc-subcomision,yardley-dinners,cnn-siestas}

To summarise the mainstream, Spanish schedules are characterised by late lunches, late dinners, late TV programs and early mornings. As a consequence the schedule yields less sleep time.

The cause of this oddity is traced back to 1940 when then ruling dictator Franco moved the clocks forward to keep in pace with neighbouring Portugal, Great Britain and France which all had undergone similar movement amidst World War II. After 1945, only France and Spain remained in the new time zone.

Does Science have a word to say? In this paper I will argue that contrary to mainstream Spanish schedules are, seventy years after the clock movement, in pace with its geographical position. That is Spaniard do the same things as Italians or Britons at (almost) the same time if geographical conditions (longitude and latitude) are considered in the problem. It is geographical coordinates and apparent motion of the Sun ---both on a daily basis and on a yearly basis--- that give sense to timetables.

For this purpose I will take use of the latest Use of Time Survey available in Spain\cite{usotiempo-2009}, Italy\cite{istat-uso-2009} and United Kingdom\cite{ukdata-timeuse-2000}, all three surveys carried out by the corresponding statistic office. By slicing the day into 144 time slots, the diary reported in these surveys allows the scanning of activities throughout the day. This study will focus on the main activities: sleeping, eating, and working. It will also analyse one location ---the place where an activity  occurs--- ``at home''. 

By scanning time slots, it is possible to track the earliest and latest occurrences of an activity or a location. Hence, one can track the wake-up time, the bed time, the first eating or breakfast, the last eating or dinner, leaving home, coming home, at work and off work. Where these labels are always related either to first or last occurrence in the day. 

Time Use surveys are poorly geographically located. All three surveys report location only at a regional level with 18 regions ---17 autonomous communities and the autonomous cities--- in Spain, 19 regions in Italy and 12 ---the 9 regions of England plus each of the remaining 3 home nations--- in United Kingdom. Data will be grouped within these regions, in fact this paper will not report data for Spain, Italy or United Kingdom but for the regions of these countries. Regions will be geographically located by computing the weighted mean latitude and longitude of all cities with a population greater than 1000 inhabitants; the weight being the population.

The surveys also trace the day of week of the reported diary journal. This paper will only analyse working day journals, avoiding week-ends. Save from this cut-off, any diary journal reporting a sleep time, an eating time, an out-of-home time slot or a working time slot will enter into the corresponding statistics.

The experimental cumulative probability of every statistics here computed shows a behaviour similar to that of the cumulative probability of a Gaussian distribution. The experimental probabilities will be then characterised by their median values. 

Figure~\ref{fig:mediana-civil} shows the median values of the distributions as a function of the mean longitude. In the figure times are given in local civil time. Regions observing WET (UTC+00) time zone ---Canary Islands in Spain and the United Kingdom--- are displayed in black. Regions observing CET (UTC+01) time zone ---remaining Spain and Italy--- are displayed in blue.\footnote{It should be here pointed out that daylight savings time is embedded in the analysis. Hence WET actually means WET or WEST depending on month of the year. In any case, what is relevant for the analysis in this paper is that these times are relative to the prime meridian or meridians shifted by $\unit{15}{\degree}$ from the prime meridian, but not to the local meridian.} In the abscissas minimum and maximum longitude of a region are shown. In the ordinate, an estimation of the error bar, based on the median absolute deviation and on the number of respondents, is also shown. In the left panel, start of day activities are shown. The end of day activities are shown in the right panel. 

\begin{figure*}[t]
  \centering
    \begin{tabular}{cc}
    \includegraphics[bb=67 82 242 312]{mediana-inicio-civil}&\includegraphics[bb=73 82 248 312]{mediana-fin-civil}\\   
  \end{tabular}
  \caption{Median values for the start-of-day activities and the end-of-day activities as a function of longitude and expressed in local civil time. Spanish times display later times as a general rule. See for instance, the bed time.}
  \label{fig:mediana-civil}
\end{figure*}

Figure~\ref{fig:mediana-civil} characterises the Spanish late time schedule. Only labour start time is similar to British and Italian counterpart.

The mainstream seems to start and end the issue by analysing these median values. However it is just the beginning of the discussion. Figure~\ref{fig:mediana-civil} really says little because there, time is not a chronological time basis but a label which depends on time zone. Blue regions are given in the CET basis, black regions are given in the WET basis. 

\begin{figure*}[t]
  \centering
    \begin{tabular}{cc}
    \includegraphics[bb=67 82 242 312]{mediana-inicio}&\includegraphics[bb=73 82 248 312]{mediana-fin}\\   
  \end{tabular}
  \caption{Median values for the start-of-day activities and the end-of-day activities as a function of longitude and expressed in WET time. The dependence with longitude is easily explained by the Sun apparent movement around the Earth which is displayed by dashed lines. Each line shows a constant solar time condition. The stripes highlights thirty minute span after and before the line.}
  \label{fig:mediana-london}
\end{figure*}

The comparison can be improved if these times are converted to a common time zone, which only implies the removal of one full hour for the regions depicted in blue. In so doing figure~\ref{fig:mediana-london} is obtained. In this figure a linear dependence between median times and longitude is revealed.

 The linear dependence can be traced by sun daily apparent motion around the Earth at a rate of four minutes per degree of longitude. This value comes from Earth's revolution period $\tau=\unit{24}{\hour}$. Dashed lines in figure~\ref{fig:mediana-london} display straight lines with slope equal to minus this quantity. Every dashed line in the figure represent a constant solar time path and the strip shows a time span of one full solar hour. ``Solar time'' measures time distance to local midnight, when the Sun crosses local nadir. Constant solar time then means equal distance to midnight or to noon. By contrast civil local time measures time distance to prime ---non-local--- meridians.

Thus figure~\ref{fig:mediana-london} shows Spanish time schedule is fairly coincidental in solar time with the British and Italian time schedule in solar time despite minor differences. %

Interestingly enough Spanish start of day median values in figure~\ref{fig:mediana-london} are always in the lower band of solar time. That means, Spanish wake-up time, breakfast time and leaving home time are a little bit earlier that the corresponding Italian or British values. This effect is stressed for the labour start time panel. In fact, if the analysis is restricted to the working people these morning differences are amplified. While this effect can be regarded as a consequence of the time zone shift from WET to CET in 1940, the following argument will show that it may also be related to natural phenomena.

The median values reported in figure~\ref{fig:mediana-london} can be corrected by longitude ---that is, converted into local solar time\footnote{Because of the issue of the daylight savings time this ``local solar time'' is not the real local solar time all the year through. Instead it is a time expressed relative to the local meridian or to the local meridian plus $\unit{15}{\degree}$ depending on the month of the year. However, the labels ``solar time'' or ``local solar time'' will be used hereafter.}--- using Earth's revolution period in the following way:
\begin{equation}
  \label{eq:1}
  T_i=t_i-\Delta_z+\langle\lambda_i\rangle\nu
\end{equation}
where $t_i$ is the median civil time value; $\Delta_{z_i}$ is the time-zone shift ---$\unit{1}{\hour}$ for the CET time zone,--- $\langle\lambda\rangle$ is the weighted mean longitude and $\nu=\tau/2\pi\sim\unit{3.82}{\hour\usk\reciprocal\radian}$ ---the constant $\nu$ is normally expressed in sexagesimal grades as four minutes per grade, as early referred.--- The label $i$ stands for a region. Eq.~(\ref{eq:1}) just rotates figure~\ref{fig:mediana-london} counterclockwise and makes the mean values expressed in solar time be roughly independent of longitude. Straightforwardly morning Spanish median solar times comes typically earlier than morning Italian and British median solar times.

Because the longitude dependence is removed after computing $T_i$ through Eq.~(\ref{eq:1}), any variability in the median solar times $T_i$ needs to be related to some other properties. The obvious choice is latitude $\phi$. 

Whereas longitude accounts for the daily apparent motion of the sun which sets the variability of noon, latitude accounts for yearly apparent motion of the sun altering solar elevation angle and mean insolation. 

Solar elevation angle is related to the so called \emph{hour angle}, latitude and solar declination by:\cite{boeker-grondelle-2011}
\begin{equation}
  \label{eq:2}
    \cos(\lambda-\lambda_s)=\frac{\sin(z)-\sin\delta_s\sin\phi}{\cos\delta_s\cos\phi}.
\end{equation}
where $z$ is the solar elevation angle above (positive) or below (negative) the horizon ---the complementary angle of the solar zenith angle---, $\phi$ is observer latitude, $\delta_s$ is the solar declination ---which is to say the latitude of the subsolar point---, $\lambda$ is observer longitude and $\lambda_s$ is the solar longitude ---the longitude of the subsolar point---. The difference $\lambda-\lambda_s$ is the hour angle. By computing $\nu(\lambda-\lambda_s)$ the hour angle is converted into time before or after local noon, and by then adding $\tau/2$ it is converted in solar time, hence its name. Notice that in Eq.~(\ref{eq:2}) $\lambda_s$ is changing on a daily basis while $\delta_s$ does so in a yearly basis. 

Eq.~(\ref{eq:2}) gets extreme values of solar time as a function of latitude for a given elevation angle $z$ when $|\delta_s|$ equals Earth's obliquity $\epsilon\sim\unit{23.5}{\degree}$ in a solstitial day. The extreme condition for the null elevation angle is then expressed by:
\begin{equation}
  \label{eq:3}
  \Delta T=\nu\cos^{-1}(-\tan(\pm\epsilon)\tan\phi)
\end{equation}
where $\Delta T$ is the time distance to local noon.

Figure~\ref{fig:latitud} shows latitude versus median values of the labour start solar time. Notice that the latitude axis is distorted: it is not $\phi$ but the inverse cosine function in Eq.~(\ref{eq:3}) that is represented in the axis. In so doing, data observing $z=0$ and $|\delta_s|=\epsilon$ in Eq.~(\ref{eq:2}) will be rendered as a straight-line in the plot. The condition $z=0$ means zero elevation angle which actually means centre of the solar discus at the horizon. The sunrise is better depicted as the instant when upper solar limbus reaches the horizon. Solar angular size and light refraction at Earth's atmosphere sets the value $z_0=-\unit{0.83}{\degree}$ as the elevation at sunrise time. In figure~\ref{fig:latitud} the solid blue line displays as a function of latitude the sunrise time for the southern solstice ($\delta_s=-\epsilon$), which is the latest possible sunrise time in the Northern hemisphere if some orbital perturbations are neglected. The coloured strip sets a time span of half an hour after or before the sunrise time.

Dashed blue lines in figure~\ref{fig:latitud} show solar elevation angles above (positive number) and below (negative number) the horizon for that day. In sharp contrast, dashed red lines show solar elevation angles for the northern solstice day ($\delta_s=\epsilon$).\footnote{Because of the the day light savings time, red lines ---corresponding to northern solstice--- have been moved accordingly one hour rightwards.} Any of these lines are extreme: the Sun is always above the condition on the right side of a blue line; the Sun is always below the condition on the left side of a red line. When a red and a blue lines intersect a point one can read the maximal and minimal solar elevation angle at the corresponding solar time and latitude.

\begin{figure*}[t]
  \centering
    \includegraphics[bb=144 65 344 268,scale=1]{latitud-inv}
  \caption{Median values of the labour start solar time as a function of latitude.  Solid blue line shows the sunrise time at the southern solstice which is the latest sunrise time of the year. The coloured strip highlights thirty minute after and before the line. Dashed blue lines set solar elevation angle conditions for this day. Red dashed lines displays solar elevation angles conditions for the northern solstice day. Selected regions are labelled. In Spain they are Canary Islands (5), Andalusia (1), Madrid (13), Castile and Le{\'o}n (7), Catalonia (9), Extremadura (11), Ceuta and Melilla (18), Basque Country (16) and Galicia (12). In United Kingdom, London (7), Scotland (11), Northern Ireland (12), Wales (10), North East (1), North West (2), West Midlands (5), South East and South West (8/9). In Italy, Lombardy (2) and Lazio (11).}
  \label{fig:latitud}
\end{figure*}

Figure~\ref{fig:latitud} shows that median values of the labour start solar time are coupled to latitude by the line of latest sunrise time. These result suggests that, statistically, this start-of-day statistic is mainly set by sunlight. The reader should notice that right of the solid blue line the Sun has already risen at any time of the year. Hence, people likes to start working after sunrise all the year through.

It is remarkable that  data and the latest sunrise time are coupled from  $\unit{28}{\degree N}$ (case of Canary Islands in Spain) to $\unit{55}{\degree N}$ (case of Scotland, Northern Ireland or the North East region of England). Canary Island is $\unit{4.5}{\degree}$ north of northern tropic, Scotland is $\unit{11}{\degree}$ south of the northern polar circle. It is reasonable to think that this dependence would be broken close to the polar circle because of the polar night.

It should be also mentioned that the yearly averaged sunrise/sunset times are independent of latitude provided that $\phi$ is not too close to polar circles. This value is set to $\uhora{06:00}$ and $\uhora{18:00}$ respectively ---the daylight savings time issue move them half an hour ahead---. Yet, latitude matters in setting social time schedules.  Also it should be stressed that the kind of dependence shown in figure~\ref{fig:latitud} can only be observed for the labour start times; end-of-day statistics are not tied to the sunset time in that way, very likely  because of the modern role of artificial light. 

Spanish timetable are revealed in Figure~\ref{fig:latitud} a little bit in the left side of the latest sunrise time line. Some issues put forward that this anomaly may not be significant:

\begin{itemize}
\item The bulk of the Spanish population lies right of the latest sunrise time. The figure displays the population by a circle whose radius is proportional to region population.
\item The time distance to sunrise is less than half an hour as observed in the figure. Which is a remarkable value if one takes into account that the time-zone scheme used worldwide is changed in steps of one hour.
\item  Solar elevation angle is always well above $\unit{-6}{\degree}$. This negative value sets the \emph{civil twilight} and within it and under clear weather conditions, illumination is sufficient for terrestrial objects to be distinguished.
\end{itemize}
More Time Use surveys should be analysed to see whether this anomaly is also observed elsewhere. That would help us to ascertain if this is a characteristic set forward by the 1940 time zone change or an adaptive response to environmental conditions.

It should be also mentioned that the figure may also be describing a competition between winter and summer conditions. Harnessing sunlight in winter mornings by approaching labour start time to the latest sunrise time line also solar elevation angle low enough in summer mornings ---with the aid of the daylight savings time,--- which may be significant below $\unit{40}{\degree}$ in latitude. In this argument it is assumed that median civil time values does not change within the year, a reasonable assumption in modern society. The daylight savings time, indeed, helps to keep morning conditions stable throughout the year, while evening conditions dramatically changes from winter to summer.\footnote{In figure~\ref{fig:latitud} the daylight savings time is set by moving red lines one hour leftwards. That is, if there were no daylight savings time, the solar elevation angle would reach$\unit{+6}{\degree}$ one hour clock earlier than displayed in the figure.}

Incidentally, figure~\ref{fig:latitud} also shows why the British 1968 time zone change was dead on arrival. That year British government moved clocks ahead from WET to CET. In figure~\ref{fig:latitud}, assuming that British time schedule has not changed too much from 1968 to 2000, that would mean shifting British data one hour leftwards. Then median values would fall well deep into the right side of the latest sunrise time, even before civil twilight. People started waking-up in winter in real night, with the solar elevation angle still below $\unit{-18}{\degree}$. Also, people started working well before sunrise in winter. The change was reversed back in 1972 amidst concerns of public safety in the activities occurring then at dawn. It should be noticed that the issue is not the problem of finding the optimal definition of the civil time. On the contrary it is, in essence, a discontinuous election where the  British government had only two choices WET time zone or CET time zone. The first choice stood and stands. 

Conversely, one can also figure out what would happen to Spanish data if the time zone were changed from CET to WET. Now data would be moved one hour rightwards. Then a time slot of solar light with little social activity would be observed. Some regions will observe that their median labour start solar time falls more than one hour after sunrise all the year through, with solar elevation angle equal to $\unit{+12}{\degree}$ even in winter. Worse, solar elevation angle at summer would lie between $\unit{30}{\degree}$ and $\unit{36}{\degree}$ at the median labour start time. Thus, labour activities with higher summer insolation, posing a potential life threat, would be promoted. Neither is there room for a midpoint solution; the Spanish government is also tied to a two-choice problem, either CET or WET. 

As final remarks, figure~\ref{fig:mediana-london} and figure~\ref{fig:latitud} show that all three surveys leads to median values which are consistent with the sun apparent motion within the interval of half a time zone. A question that arises is why Spanish timetable are coordinated in such a way with Italy and United Kingdom despite using a time zone which is ahead from its mean meridian. The answer to that question is quite simple from a scientific point of view: assuming Spanish clocks where synchronised with sun in 1940, the clock movement would have broken this harmony and should have triggered a social relaxation process which moved timetable back in sync with Sun within some relaxation time. Indeed it may have taken little time to bring timetables back to sync due to the special circumstances occurring in 1940, with the country devastated after a civil war. The kind of circumstances which failed to occur when United Kingdom unsuccessfully mimicked the movement in 1968. 

As its always the case, the relaxation process should minimise the alteration. Because the alteration moved clocks forward, the process moved time tables backwards to late ``by the clock'' times. The success of the relaxation process in Spain is acknowledge by figure~\ref{fig:mediana-london} and figure~\ref{fig:latitud} seventy years after the change. Also indirectly the long-term stability of the change acknowledges the existence of this process.

Summarising, Spanish timetable looks remarkably coincident with Italian and British timetable if longitude is taken into account ---see figure~\ref{fig:mediana-london}--- and if latitude is taken into account ---see figure~\ref{fig:latitud}.--- All three statistics reports data which looks like tightly coupled to sun apparent motion.

Social time schedule then looks like related to sun light, specially in the start of day (morning) regime. The actual mess of the Spanish timetable discussion can be best summarise in the following quotation of the Spanish Legislature when addressing the idea of changing time zone back to WET: 
\begin{quote}
  ``Our timetable is determined more by the sun than by the clock. We eat at one o'clock in the afternoon and dine at eight, according to the sun, but the clock says it is three o'clock and 10 o'clock''\cite{informe-subcomision,bbc-subcomision}
\end{quote}
where the Branch arguably fails to acknowledge that it is the Sun, not clocks, that still sets timetables in the 21st century.

\acknowledgements

The author wishes to thank Instituto Nacional de Estad{\'\i}stica INE, Istituto Nazionale di Statistica Istat, and UK Data Service, for providing access the Time Use survey microdata.